\documentclass[journal]{IEEEtran}%
\usepackage{amsfonts}
\usepackage{amsmath}
\usepackage{amssymb}
\usepackage{amsthm}
\usepackage{graphicx}%
\usepackage{color}
\providecommand{\U}[1]{\protect\rule{.1in}{.1in}}
\newtheorem{theorem}{Theorem}

\newtheorem{corollary}{Corollary}

\newtheorem{definition}{Definition}

\newtheorem{remark}{Remark}

\begin{document}

\title{Enabling Data Exchange in Interactive State Estimation under Privacy Constraints}
\author{E.V. Belmega, \emph{Member, IEEE,} L. Sankar, \emph{Member, IEEE,} and H. V.
Poor, \emph{Fellow, IEEE} \thanks{The material in this paper was partially presented  at the IEEE Intl. Conf. on Network Games, Control and Optimization
(NETGCOOP), Avignon, France, Nov. 2012 \cite{belmega-netgcoop-2012}. This
research was supported by the National Science Foundation under Grants
DMS-118605 and CCF-1016671. \newline E.V. Belmega is with ETIS/ENSEA-UCP-CNRS,
Cergy-Pontoise, France, belmega@ensea.fr; L. Sankar is with Arizona State
University, Tempe, AZ, USA, lalithasankar@asu.edu; and H.V. Poor is with Dept.
of Electrical Engineering, Princeton University, Princeton, NJ, USA,
poor@princeton.edu. } }
\maketitle

\begin{abstract}
Data collecting agents in large networks, such as the electric power system, need to share information (measurements) for estimating the system state in a distributed manner. However, privacy concerns may limit or prevent this exchange leading to a tradeoff between state estimation fidelity and privacy (referred to as \textit{competitive privacy}). This paper builds upon a recent information-theoretic result (using mutual information to measure privacy and mean-squared error to measure fidelity) that quantifies the region of achievable distortion-leakage tuples in a two-agent network. The objective of this paper is to study centralized and decentralized mechanisms that can enable and sustain non-trivial data exchanges among the agents.  
A centralized mechanism determines the data sharing policies that optimize a network-wide objective function combining the fidelities and leakages at both agents. Using common-goal games and best-response analysis, the optimal policies allow for distributed implementation. In contrast, in the decentralized setting, repeated discounted games are shown to naturally enable data exchange without any central control nor economic incentives. The effect of repetition is modeled by a time-averaged payoff function at each agent which combines its fidelity and leakage at each interaction stage. For both approaches, it is shown that non-trivial data exchange can be sustained for specific fidelity ranges even when privacy is a limiting factor.
\end{abstract}


\vspace{-0.1in}

\section{Introduction}

\label{sec:intro}
The increasing demand for sustainable energy in the information era requires a highly efficient and reliable electric power system in which renewables can be effectively integrated. Given the size and complexity of the electric network, sustained and reliable operations involve an intelligent cyber layer that enables distributed monitoring, processing and control of the network. In fact, data collection and processing is performed locally at various collecting entities (e.g., utility companies, systems operators, etc.) that are spread out geographically. The interconnectedness of the network requires that these distributed entities share data amongst themselves to ensure precise estimation and control, and in turn, system stability and reliability. Despite its importance, data sharing in the electric power system is limited - sometimes with catastrophic consequences \cite{FERC,FERC_WAM} - because of competitive interests or privacy concerns. Furthermore, this problem of distributed computation, control and data sharing is not specific to electrical power networks and may arise in other critical infrastructure networks (e.g., air transport, electronic healthcare, and the Internet). We henceforth refer to this problem as competitive privacy as in \cite{sankar-sgcom-2011}.

The notion of privacy is predominantly associated with the problem of ensuring that personal data about individuals, stored in a variety of databases or cloud servers, is not revealed. Quantifying the privacy of released data has captured a lot of attention from the computer science and information theoretic research communities leading to two different rigorous frameworks: differential privacy introduced by Dwork et al. \cite{Dwork06differentialprivacy}, \cite{Dwork_DP_Survey}; and information-theoretic privacy developed in \cite{Sankar_ITPrivacy}. The first framework focuses on worst-case guarantees and ignores the statistics of the data; while the latter focuses on average guarantees and is cognizant of the input data statistics; their appropriateness depends on the application at hand. In the information era, however, privacy restrictions also appear in data exchange contexts as detailed here; it was first studied via an information-theoretic framework in \cite{sankar-sgcom-2011}.

For the distributed state estimation problems via data exchange (as applied to the electric power system), the information-theoretic competitive privacy framework holds the following advantages: (a) takes into account the statistical nature of the measurements and underlying state (e.g., complex voltage measurements in the grid that are often assumed to be Gaussian distributed); (b) combines both compression and privacy in one analysis by developing rate and privacy optimal data sharing protocols; and (c) quantifies privacy over all possible sequence of measurements and system states. 

The competitive privacy information-theoretic framework introduced recently in \cite{sankar-sgcom-2011} studies data sharing among two interconnected agents when privacy concerns limits data sharing, and therefore, the fidelity of distributed state estimation performed by the agents. The authors proposed a distributed source coding model to quantify the information-theoretical tradeoff between estimate fidelity (distortion via mean-square error), privacy (information leakage), and communication rate (data sharing rate). Every achievable distortion-leakage tuple represents a four-dimensional vector of opposing quantities that cannot be optimal simultaneously; \emph{minimum distortion} for one agent implies \emph{maximum leakage} for the other; \emph{minimum leakage} for one agent implies \emph{maximum distortion} for the other. A pertinent question follows: how to choose such a tradeoff in practice?

The objective of this work is to address this question via mechanisms that can enable and sustain specific distortion-leakage tuples in both centralized (a unique decision-maker) and decentralized settings (each agent has his own individual agenda). Game theory is a mathematical toolbox for studying interactions among strategic agents and has established its value in a wide-variety of fields including wireless communications \cite{lasaulce-book-2011}, \cite{han-book-2012}. While often applied in the non-cooperative and decentralized context, even in centralized settings, game theory can be valuable when devising efficient and distributed algorithms to compute the solution; in fact, these tools can be very useful to solve difficult, non-convex problems that arise in multi-agent models with multiple performance criteria (such as leakage vs. fidelity) as we present later in the sequel.

Our first approach assumes a central controller that imposes the data sharing choices of the two agents (e.g., when electric utility companies share their data with a central systems operator). The network-wide objective function captures both, the overall leakage of information and the total distortion of the estimates of the two agents via their weighted sum. To circumvent the non-convexity of this objective function, we exploit the parallel between \emph{distributed optimization} problems and \emph{potential games} \cite{monderer-eco-1996}. The Nash equilibria of the resulting common-goal game are the intersection points of the best-response functions which turn out to be piece-wise affine. Moreover, using game theoretic tools we provide a distributed algorithm - the iterative best-response algorithm - that converges to an optimal solution. Our results show that the central controller can smoothly manipulate the distortion-leakage tradeoff between two extremes: both users share fully their data (\emph{minimum distortion - maximum leakage}) or not at all (\emph{maximum distortion - minimum leakage}). Specifically, not all information-theoretic tuples can arise as outcomes, but only the optimizers of the network-wide objective function.

If there is no central controller (e.g., when agents are two systems operators that need to share data to monitor large parts of the electric grid), each agent chooses its own data sharing strategy to optimize its individual distortion-leakage tradeoff. In \cite{belmega-isccsp-2012}, we showed that data sharing decreases the distortion of the agent receiving data while the sharing agent only increases its leakage. Thus, when the interaction takes place only once (i.e., one-shot interaction), rational agents have no incentive to share data. Economic incentives overcome this issue \cite{belmega-isccsp-2012} and all distortion-leakage tuples can be achieved assuming that agents are paid (by a common moderator) for their information leakage. 

In the second part of this paper, we show that pricing is not the only mechanism enabling cooperation. If the agents interact repeatedly over an indeterminate period, \emph{tit-for-tat} type of strategies (i.e., an agent shares his data \emph{as long as} the other agent does the same) turn out to be stable outcomes of the new game. We show that a whole sub-region of distortion-leakage tuples (in between the aforementioned extremes) is achieved without the need for a central authority; effectively, the agents build trust by exchanging data in the long term. 

Preliminary results regarding the repeated interaction have been presented in \cite{belmega-netgcoop-2012}. We provide here a complete analysis and detailed proofs. Moreover, in this current version, we: (i) introduce different discount factors to model individual preferences for present vs. future rewards; (ii) give closed-form bounds on the discount factors; and (iii) illustrate more results. 

The paper is organized as follows. In Section \ref{sec:sys}, we
introduce the system model and an overview of the most relevant information and game-theoretic concepts and prior results. The common goal non-cooperative game and its Nash equilibria are analysed in Section \ref{sec:potential_game} as a simpler alternative to a non-convex centralized. In Section \ref{sec:repeated_game} in which we show that repetition the repeated games framework and study its solutions and achievable distortion-leakage pairs. Numerical results that illustrate the analysis are also provided. We conclude in Section \ref{sec:clz}.


\vspace{-0.1in}

\section{\label{sec:sys}System Model}

\begin{center}
\begin{figure}
\caption{{\protect\scriptsize {Network of physically interconnected nodes. We focus on two communicating nodes/agents and exploit the possibility of exchanging information about their local measurements. }}}%
\label{fig:sys}%
\includegraphics[width=0.85\columnwidth]{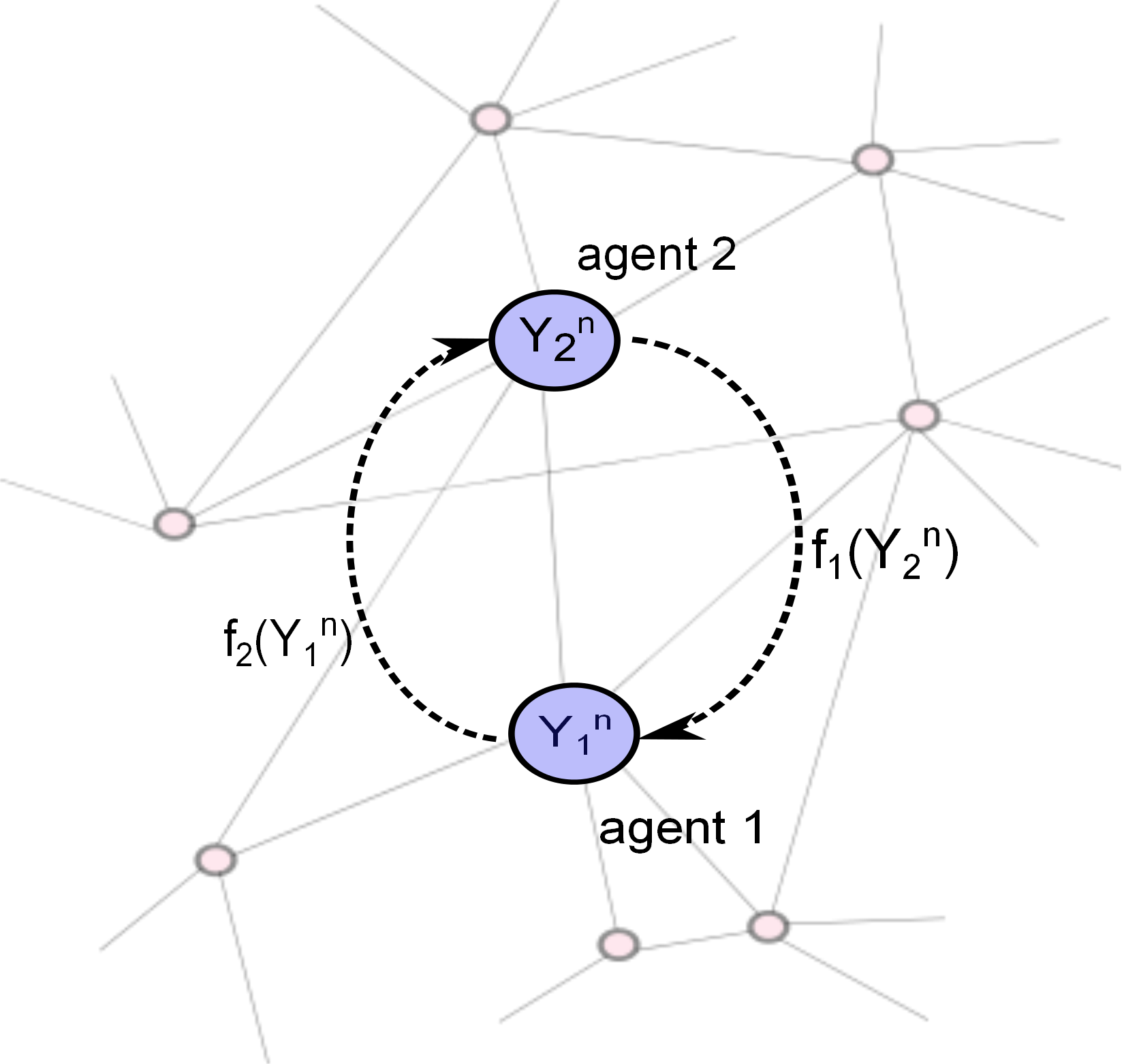} \vspace
{-0.2in}\end{figure}
\end{center}

We consider a network composed of physically interconnected nodes as illustrated in Figure \ref{fig:sys}. We focus only on a pair of such nodes - called agents - which are capable of communicating and sharing some of their collected data.  
\vspace{-0.2in}Each
agent observes a sequence of $n$ measurements from which it estimates a set of system parameters, henceforth referred to as states. 
The measurements at each agent are also affected by the states of the other agent. For simplicity reasons, we consider a linear approximation model (e.g., model of voltages in the electric power network \cite{AburBook}). Denoting the state and measurement vectors at agent $j\in\{1,2\}$ as $X_{j}^{n}$ and
$Y_{j}^{n}$, respectively, the linear model is:%
\begin{equation}%
\begin{array}
[c]{lcl}%
Y_{1,k} & = & X_{1,k}+\alpha_1 X_{2,k}+Z_{1,k}\\
Y_{2,k} & = & \alpha_2 X_{1,k}+X_{2,k}+Z_{2,k},
\end{array}
\label{sys_model}%
\end{equation}
where $\alpha_1$, $\alpha_2$ are positive parameters. The $k^{th}$ states, $X_{1,k}$ and $X_{2,k}$,
for all $k,$ are assumed to be independent and identically distributed
(i.i.d.) zero-mean unit-variance Gaussian random variables and the additive
zero-mean Gaussian noise variables, $Z_{1,k}$ and $Z_{2,k}$, are assumed to be
independent of the agent states and of fixed variances $\sigma_{1}^{2}$ and
$\sigma_{2}^{2}$, respectively.

This model is relevant to direct current (DC) state estimation problems in which the agents (e.g., system operators or energy management entities) need to share their local measurements (e.g., power flow and injections at specific locations) to estimate with high fidelity their local states (e.g., complex voltages).

Agent $j$ can improve the fidelity of its state estimate if the other agent $i \neq j$ decides to share some information regarding his measurements - say $f_{j}\left(  Y_{i}^{n}\right)$. At the same time, the amount of agent $i$ leakage on his state information is constrained (in the competitive privacy framework of \cite{sankar-sgcom-2011}). These conflicting aspects are measured by information-theoretic concepts: the desired
fidelity and privacy amount to meeting a distortion (mean-squared error) and a information leakage constraint, respectively:
\begin{subequations}
\label{metrics}%
\begin{align}
\mathbb{E}\left[  \frac{1}{n}\sum_{k=1}^{n}\left(  X_{j,k}-\hat{X}%
_{j,k}\right)^2  \right]   &  \leq D_{j},\text{ and}\\
\frac{1}{n}I(X_{j}^{n};f_{i}(Y_{j}^{n}),Y_{i}^{n})  &  \leq L_{j},
\label{eq:leakage}%
\end{align}
\end{subequations}
where $D_{j}$ represents the distortion of estimate $\hat{X}_{j}^{n}$ - which depends on the other agent's sharing policy, $f_j(Y_i^n)$ - from the actual state $X_{j}^{n}$, and $L_{j}$ is the maximum information leakage. The mutual information in (\ref{eq:leakage}) measures the average leak of information per sample about the private state $X_{j}^{n}$ of agent $j$ to the other agent. The other agent can infer information on $X_{j}^{n}$ from two sources: (i) his own measurements $Y_i^n$ 
(\ref{sys_model}); and (ii) the data shared by agent $j$, i.e., $f_{i}\left(  Y_{j}^{n} \right)$.

Sankar et al. \cite{sankar-sgcom-2011} determined the entire region of achievable 
$\left(  D_{1},L_{1},D_{2},L_{2}\right)$ tuples. The authors devised a particular coding scheme - based on quantization and binning techniques - that satisfies the distortion constraints $D_{i}$ and achieves the minimal leakage $L_{j}$ constraint (for both agents). We summarize the resulting achievable distortion-leakage (DL) region in the following theorem. 

\begin{theorem}
\label{Theorem1} \cite{sankar-sgcom-2011} The
distortion-leakage tradeoff for a two-agent competitive privacy problem (described above) is the four-dimensional set of all $\left(  D_{1},D_{2},L_{1},L_{2}\right)  $
tuples such that: \\
For all $i,j\in\{1,2\},$ $i\neq j$, 
\begin{itemize}
\item $D_{j} < D_{\max ,j}$ \ : 
\begin{align}
L_{i}(D_j)& = \frac{1}{2}\log \left( \frac{m_{i}^{2}}{m_{i}^{2}D_{\min
,i}+n_{i}^{2}(D_{j}-D_{\min ,j})}\right) ;  \label{Th_L1}
\end{align}%
\item $D_{j} \geq D_{\max ,j}$ : $L_{i} (D_j) = \log \left( V_{j}/(V_{j}-\alpha_{j})\right)/2$,
\end{itemize}
with the parameters $V_{j}=1+\alpha_j ^{2}+\sigma _{j}^{2}$, \ $E=\alpha_1 + \alpha_2$,  \ $n_{j}=(V_{i}-\alpha_i E)/(V_{1}V_{2}-E^{2})$, \ $m_{j}=(\alpha_j V_{i}-E)/(V_{1}V_{2}-E^{2})$, \ and 
\begin{subequations}
\begin{align}
\nonumber
D_{\min ,j}& =1-\frac{(\alpha_i^{2}V_{j}+V_{i}-2\alpha_i  E)}{(V_{1}V_{2}-E^{2})}, \\ \nonumber
D_{\max ,j}& =1-1/V_{j}.
\end{align}%
\end{subequations}
\end{theorem}
The maximal and minimal distortions, denoted by $D_{\max,j}$ and $D_{\min,j}$, represent the extreme cases in which, the other agent $i$, either sends no information or fully discloses his measurements. If $D_{j} < D_{\max ,j}$, the distortion constraint is non-trivial and agent $i$ has to leak information about his own state. The leakage is increasing with $D_j$. If $D_{j} \geq D_{\max ,j}$, the distortion constraint is trivial, and agent $i$ does not have to send any data. His minimum leakage is not zero because agent $j$ can still infer some private data (on agent $i$ state) from his measurements $Y_j^n$. 

Notice that the region contains asymmetric tuples in terms of data sharing. This results from the opposing distortion and leakage components that cannot be optimal simultaneously: \emph{minimum distortion} at one agent corresponds to \emph{maximum leakage} at the other, and \emph{minimum leakage} of one agent corresponds to \emph{maximum distortion} at the other. From this region (which is four dimensional) alone, it is not clear how to choose such a tradeoff tuple. In this paper, the main objective is to study different mechanisms that explain how specific tuples may arise in centralized and decentralized settings.


\vspace{-0.1in}

\section{Centralized solution via common goal games}
\label{sec:potential_game}

Reliability in the North American electric power network is ensured by regulatory bodies (such as the North American Electric Regulatory Corporation (NERC) \cite{NERC}, Federal Electricity Regulatory
Commission (FERC)\ \cite{FERC}) and enforced by regional and independent system operators. Our first approach is focused on centralized networks in which a central controller dictates the data-sharing policies of the two agents.

The controller wishes to minimize both the overall estimation fidelity and the information leakage. But, as discussed in the previous section, the two objectives are opposing and they cannot be optimized simultaneously; a network-wide compromise has to be made.

In multi-objective optimization problems, scalarization via the weighted sum of the different objectives is a common technique that provides good tradeoff tuples by solving a simpler scalar problem instead. In some cases (such as convex optimization problems), the tuples obtained by tuning the weights among the objectives are all optimal tradeoffs \cite{boyd-book-2004}.

The network-wide objective function that captures the tradeoff between overall estimation fidelity and leakage - by their weighted sum - writes as follows: 
\begin{equation}
{\small u_{\mathrm{sys}}(D_{1},D_{2})= -\sum_{j=1}^{2}L_{j}(D_{i})+\frac{q}%
{2}\log\left(  \displaystyle{\sum_{j=1}^{2}}\overline{D}_{j} \left/
\displaystyle{\sum_{j=1}^{2}}D_{j}\right.  \right)  , }
\nonumber\label{eq:system_wide}%
\end{equation}
where the leakage of information $L_j(D_i)$ is given 
in \eqref{eq:leakage} and $q = \hat{w}/\tilde{w}>0$ is the ratio of the weighting factors between the two terms. 

For homogeneity reasons, the second term has to relate to logarithmic information measures. We propose to balance the information leakage (in bits/sample) with the overall shared information (also in bits/sample) which is inversely proportional to the distortion \cite[Chap. 10]{cover-book-2006}; as the distortions decrease, the information revealed per sample (or communication rate) increase. 


The problem reduces to finding the distortion pairs $(D_1,D_2)$ - characterizing the data-sharing policies of both users - which maximize the objective function \eqref{eq:system_wide}. One can easily check that this function is not always concave on its domain. By using a distributed approach to find the solution, we can overcome this obstacle. Assume each agent controls his own data-sharing policy which impacts directly on the distortion at the other agent. The control parameter (or action) of agent $j$ is denoted by $a_j = D_i$. The agents choices are driven by the same \emph{common goal}, i.e., the network-wide objective function. 

We further exploit the parallel between distributed optimization and \emph{potential games} which has several advantages: (i) allows to solve a non-convex problem in a simpler manner; (ii) leads to an iterative and distributed procedure that converges to a local optimal tradeoff tuple; and (iii) the central controller can manipulate this outcome by tuning a scalar parameter alone. The partial shift of intelligence, from the centralized controller towards the agents, paves the way of developing scalable data-sharing policies in more complex networks (of large number of communicating agents).

We model the \emph{common goal game} by $\mathcal{G}_{\mathrm{sys}%
}=(\mathcal{P},\{\mathcal{A}_{j}\}_{j\in\mathcal{P}},u_{\mathrm{sys}})$ in which
$\mathcal{P}\triangleq\{1,2\}$ designates the set of \emph{players} (the two agents);
$\mathcal{A}_{j}$ is the set of actions that agent $j$ can take. The payoff function of both players, 
$u_{\mathrm{sys}} : \mathcal{A}_{1} \times\mathcal{A}_{2} \rightarrow
\mathbb{R}$, is given by%
\begin{equation}
{\small u_{\mathrm{sys}}(a_{1},a_{2})= -\sum_{j=1}^{2}L_{j}(a_{j})+\frac{q}%
{2}\log\left(  \displaystyle{\sum_{j=1}^{2}}\overline{D}_{j} \left/
\displaystyle{\sum_{j=1}^{2}}a_{j}\right.  \right)  , }
\nonumber\label{eq:potential_gen}%
\end{equation}
The utility
function can be re-written using Theorem \ref{Theorem1} as%
\begin{equation}
{\small u_{\mathrm{sys}}(a_{1},a_{2})=\frac{1}{2}\log\left(  \frac{(\gamma
_{1}a_{1}+\delta_{1})(\gamma_{2}a_{2}+\delta_{2})}{(a_{1}+a_{2})^{q}}\right)
+ C_{0}, } \label{eq:potential}%
\end{equation}
where $\gamma_{j}=(n_{j}/m_{j})^{2}$ and $\delta_{j}=D_{\min,j}-\gamma
_{j}D_{\min,i}$, $C_{0}=q/2\ \log(\overline{D}_{1}+\overline{D}_{2})$. Without
loss of generality, the additive constant $C_{0}$ and the multiplicative
positive constant $1/2$ in the payoff function can be ignored in the following
analysis of the NE \cite{weibull-book-1997}.

The non-cooperative game $\mathcal{G}_{\mathrm{sys}}$ falls into a special class called \emph{potential games} \cite{monderer-eco-1996} that have many
interesting properties. Their particularity lies in the existence of a global function -
called \emph{potential function} - that captures the players'
incentives to change their actions. In our case, the network-wide objective 
\eqref{eq:potential} represents precisely the \emph{potential
function} of the game. Monderer et al. \cite{monderer-eco-1996} proved that
every potential game has at least one Nash Equilibrium (NE) solution\footnote{Nash equilibrium represents the natural solution concept in non-cooperative games \cite{tirole-book-1991} defined as a profile of actions (one action for each agent) which is stable to unilateral deviations. Intuitively, if the players are at the NE, no player has any incentive to deviate and switch its action unilaterally (otherwise, the deviator decreases its payoff value).}. Also, every local maximizer of the potential is NE of the game. However, since the
potential function is not concave \cite{neyman-gt-1997}, the game
may have other NE points (e.g., certain saddle points of the potential function).

To completely characterize the set of all NE, we study of the best-response
correspondence defined by:
\begin{align*}
&\begin{array}[c]{lcl}%
BR_{j} : \mathcal{A}_{i} \rightarrow\mathcal{A}_{j} &  & \\
\text{s.t. }BR_{j}(a_{i}) = \arg\sup_{b_{j}}u_{\mathrm{sys}}(b_{j},a_{i}), &
&\\
&  &
\end{array} & \\
&\begin{array}
[c]{lcl}%
 BR : \mathcal{A}_{1} \times\mathcal{A}_{2} \rightarrow
\mathcal{A}_{1} \times\mathcal{A}_{2} &  & \\
\text{s.t. }BR(a_{1},a_{2}) = (BR_{1}(a_{2}) \times BR_{2}(a_{1})). &  &
\end{array}& \label{eq:BR}
\end{align*}
The best-response (BR) of agent $j$ to an action $a_i$ played by the other agent $i$ - denoted by $BR_{j}(a_i)$ -
is the optimal choice (payoff maximizing one) of agent $j$ given the action of the other player. The best-response correspondence, $BR(\cdot, \cdot)$, 
represents the concatenation of both agents' BRs. The optimal action of agent $j$ for fixed choices of the other agents might not be a singleton, hence the correspondence
definition (a set-valued function).

Nash \cite{nash-academy-50} showed that the fixed points of the BR
correspondence are the NE. In our case, the BR functions reduce to simply piecewise affine functions. Thus, the game $\mathcal{G}_{\mathrm{sys}}$ can be described as a \emph{``Cournot duopoly''} interaction \cite{tirole-book-1991} in which 
the set of NE points is completely characterized by intersection points of the best-response functions $BR_{1}(\cdot)$ and $BR_{2}(\cdot)$ \cite{belmega-eurasip-2010}. Using game theoretical tools, we reduce the non-convex optimization problem to the analysis of intersection points of piecewise affine functions.

We further investigate a refined stability property of NE, namely,
their \emph{asymptotic stability} \cite{tirole-book-1991}. This property is
important when the game has multiple NE. In such cases it seems \emph{a priori} impossible to predict which particular NE will be the actual outcome. Nevertheless, if the players update their choices using the best-response dynamics - the
agents sequentially choose their best-response actions to previously observed
plays by the others  \cite{lasaulce-book-2011}) - the outcome of a 
``Cournot duopoly'' can be predicted
exactly, depending on the initial point. To be precise, the asymptotic stable NE will be the attractors of this dynamics whereas the other NE will not be observed generically (except when the initial point
happens to be one of these NE). For a more detailed discussion
on ``Cournot duopoly'' the reader is referred to \cite{tirole-book-1991}, \cite{belmega-eurasip-2010}.

To compute the BRs, we analyze the first-order partial derivatives of the potential function. We
distinguish different behaviors depending on the emphasis on either the leakage
of information ($q\leq 1$) or estimation fidelity ($q > 1$). 

\vspace{-0.15in}

\subsection{Emphasis on the fidelity of state estimation ($q>1$)}

\label{subsec:fidelity}

By developing the first-order partial derivatives of the potential function,
the best-responses become:%
\begin{equation}
\label{eq:br_fidelity}BR_{j}(a_{i}) = \left\{
\begin{array}
[c]{ll}%
F_{j}(a_{i}), & \text{if} \ \ D_{\min,i} < F_{j}(a_{i}) \leq\overline{D}%
_{i},\\
\overline{D}_{i}, & \text{if} \ \ F_{j}(a_{i}) > \overline{D}_{i},\\
D_{\min,i} & \text{otherwise},
\end{array}
\right.
\end{equation}
where $F_{j}(a_{i}) = a_{i}/(q-1) - q \delta_{j}/((q-1)\gamma_{j})$ is an
affine function of $a_{i}$ with parameters $\gamma_{j} = (n_{j}/m_{j})^{2}$, $\delta_{j}
= D_{\min,j} - \gamma_{j} D_{\min,i}$. The intersection point of the two affine functions $F_{1}(\cdot)$ and $F_{2}(\cdot)$ is

\begin{equation}
\label{eq:intersection}
\left\{ \begin{array}{lcl}
a_1^* & = & \frac{q}{1-(q-1)^2} \left( \frac{\delta_1}{\gamma_1} (q-1) + \frac{\delta_2}{\gamma_2} \right)\\
a_2^* & = & \frac{q}{1-(q-1)^2} \left( \frac{\delta_2}{\gamma_2} (q-1) + \frac{\delta_1}{\gamma_1} \right).
\end{array} \right.
\end{equation}

The NE can be completely characterized by
the intersection points of the two BR functions in the profile set, i.e.,
$\Delta\triangleq[D_{\min,2}, \overline{D}_{2}]\times[D_{\min,1}, \overline
{D}_{1}]$. Noticing that the BRs are piecewise affine functions, the following result is obtained.
\begin{theorem}
\label{Theorem2}
The game $\mathcal{G}_{\mathrm{sys}}$ has generically a unique or three NE assuming the central controller puts an emphasis on the overall state estimation fidelity, i.e., $q>1$. In very specific cases (on the system parameters), the game may have an infinite number of NE (when the affine functions $F_j(\cdot)$ are identical) or two NE (when the intersection point \eqref{eq:intersection} lies on the border of $\Delta$).
\end{theorem}

Intuitively, if the network parameters $\alpha_1$ and $\alpha_2$ are randomly drawn from a continuous distribution, the probability of having an infinite or two NE is zero. In general, depending on the relative slopes of the two BRs, the game has a unique NE (given by \eqref{eq:intersection} provided it lies in $\Delta$) or three NE (one is \eqref{eq:intersection} and the other two lie on the border of $\Delta$). The details of the proof are given in Appendix \ref{appendix:A}. In this case, the NE of the common goal game are either network-wide optimal or saddle points of the central controller's objective function (also the potential function of the game). However, only the NE that are optimizers of this objective function are asymptotically stable and can be observed as outcomes of best-response dynamics/algorithms.
 
\vspace{-0.1in}

\subsection{Emphasis on the overall leakage of information ($q \leq 1$)}
\label{subsec:equal}

As opposed to the previous case, the BR of agent $j$ is a piecewise constant function given as follows:%
\begin{equation}
\label{eq:br_leakage}BR_{j}(a_{i}) = \left\{
\begin{array}
[c]{ll}%
\overline{D}_{i}, & \text{if} \ \ {\bf{C_i^{[q=1]}}} \ \text{or} \ {\bf{C_i^{[q<1]}}} \\
D_{\min,i} & \text{otherwise},
\end{array}
\right.
\end{equation}
with the following conditions
\begin{align*}
&\nonumber {\bf{C_i^{[q=1]}}}: \ \ q = 1 \ \text{and} \ a_{i} > \frac{\delta_{i}}{\gamma_{i}},&\\
\nonumber
&{\bf{C_i^{[q<1]}}}: \ \ q < 1 \ \text{and} \ F_{j}(a_{i}) > \overline{D}_{i} \text{ or
}&\\
\nonumber
&\left(  D_{\min,i} < F_{j}(a_{i}) \leq\overline{D}_{i} \ \text{ and }
\right. \left.  u_{\mathrm{sys}}(D_{\min,i}, a_{i}) \leq u_{\mathrm{sys}}
(\overline{D}_{i}, a_{i}) \right),&  
\end{align*}
where $F_{j}(a_{i})$ is defined in (\ref{eq:br_fidelity}). The intersection points of such functions switching between the two extremes, can only lie on the corner points of $\Delta$.

\begin{theorem}
\label{Theorem4} The game $\mathcal{G}_{\mathrm{sys}}$ has either a unique or two NE assuming the central controller puts an emphasis on the leakage of information ($q \leq 1$). The NE lie on the four corners of $\Delta$, depending on the system parameters.
\end{theorem}
When the game has two NE, they are always given by the two symmetric extreme corners $(D_{\min,2}, D_{\min,1})$ (both users fully disclose their measurements) and $(\overline{D}_{2}, \overline{D}_{1})$ (no cooperation). Otherwise, either of the four corners can be the outcome of the game, depending on the system parameters. Also, all NE are asymptotically stable in this case. The proof is omitted as it is tedious and follows simply by analysing the intersection of piecewise constant functions. 
In this case, the central controller cannot smoothly manipulate the outcome by tunning $q \in [0, 1]$ and only extreme distortion-leakage pairs are achieved. In the remainder of this section, we focus only on the case of $q > 1$, the controller puts an emphasis on the estimation fidelity.
\vspace{-0.1in}

\subsection{Numerical results}

\label{subsec:simus_potential}

\begin{center}
\begin{figure}[h!]
\includegraphics[width=.9\columnwidth]{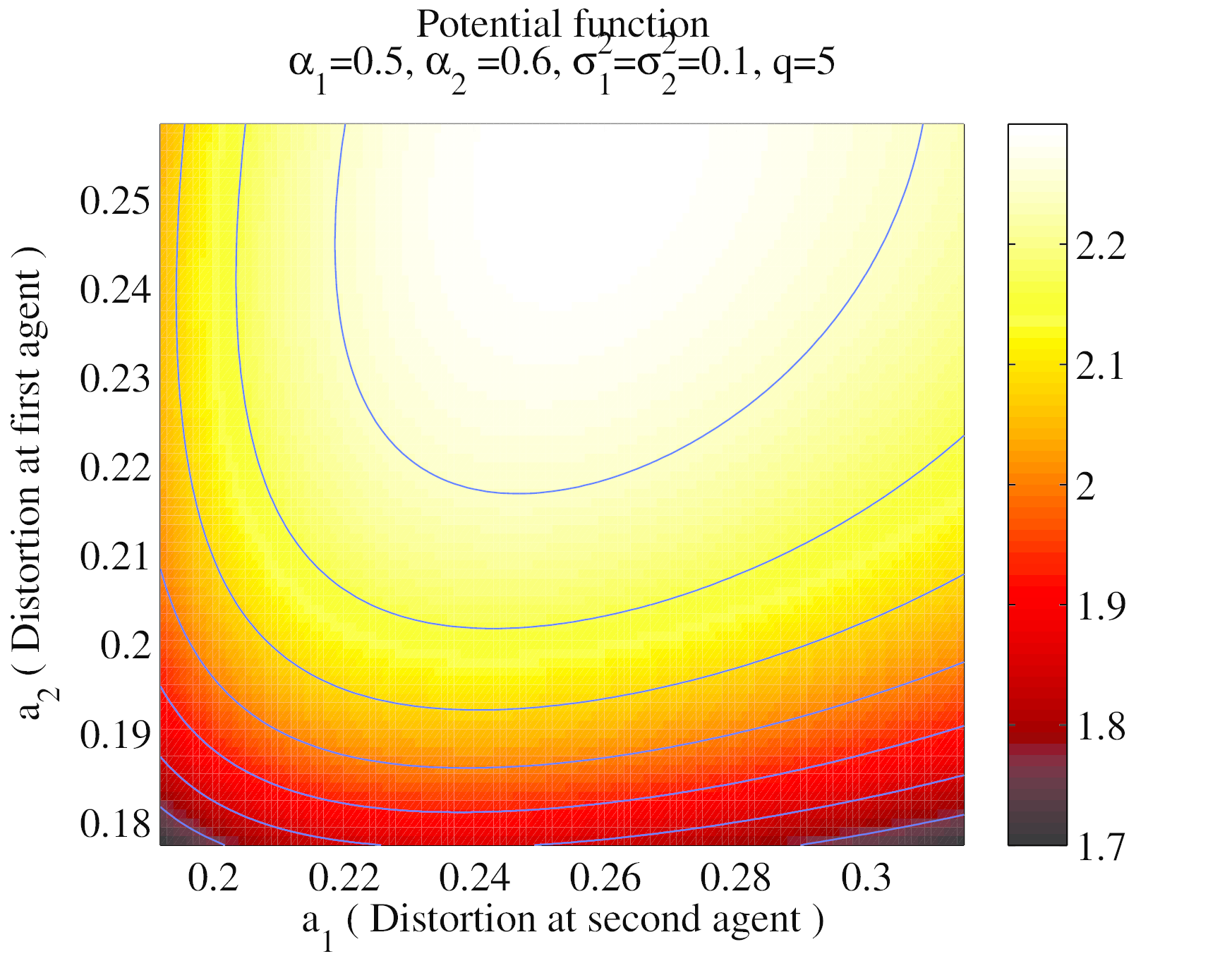}
\includegraphics[width=.9\columnwidth]{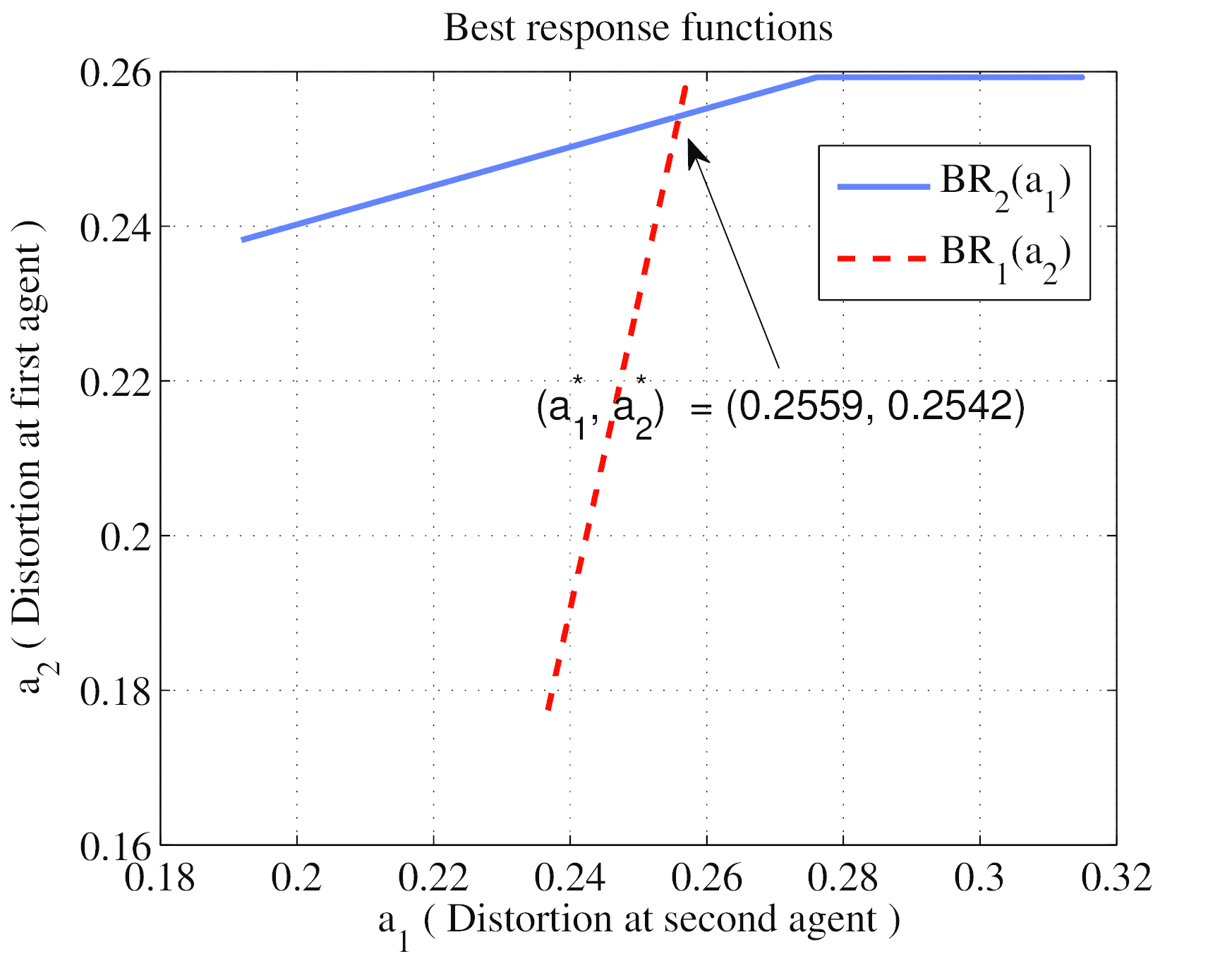} 
\caption{{\protect\scriptsize {Water-levels of the potential (up) and
BRs (down) as functions of $(a_{1},a_{2}) \in(D_{\min,2}, \overline{D}_{2}]
\times(D_{min,1}, \overline{D}_{1}]$. The potential has a unique maximum
equal to $(a_{1}^{*}, a_{2}^{*})$ which is the asymptotically stable NE.}}}%
\label{fig:fig1_potential_1NE}%
\end{figure}
\end{center}

We assume the target distortions to be equal to the maximum distortions $\overline{D}_{j} = D_{\max,j}$, $j \in\{1,2\}$. First, we consider the case in which a unique NE exists and $q > 2$.
Fig. \ref{fig:fig1_potential_1NE} illustrates the water-levels of the
potential function and the BRs in $\Delta$ for the
scenario: $\alpha_1 = 0.5$, $\alpha_2= 0.6$ and $\sigma_{1}^{2} = \sigma_{2}^{2} =
0.1$. The NE is the intersection point $(a_{1}^{\mathrm{NE}}, a_{2}^{NE}) =
(a_{1}^{*}, a_{2}^{*}) = (0.2559, 0.2542)$ and is asymptotically stable. Using a best-response iteration, the two agents converge always - from any initial point - to the optimal point. If a small perturbation occurs, using the same iterative BR dynamics, the agents will return to this point.

\begin{center}
\begin{figure}[h!]
\includegraphics[width=.9\columnwidth]{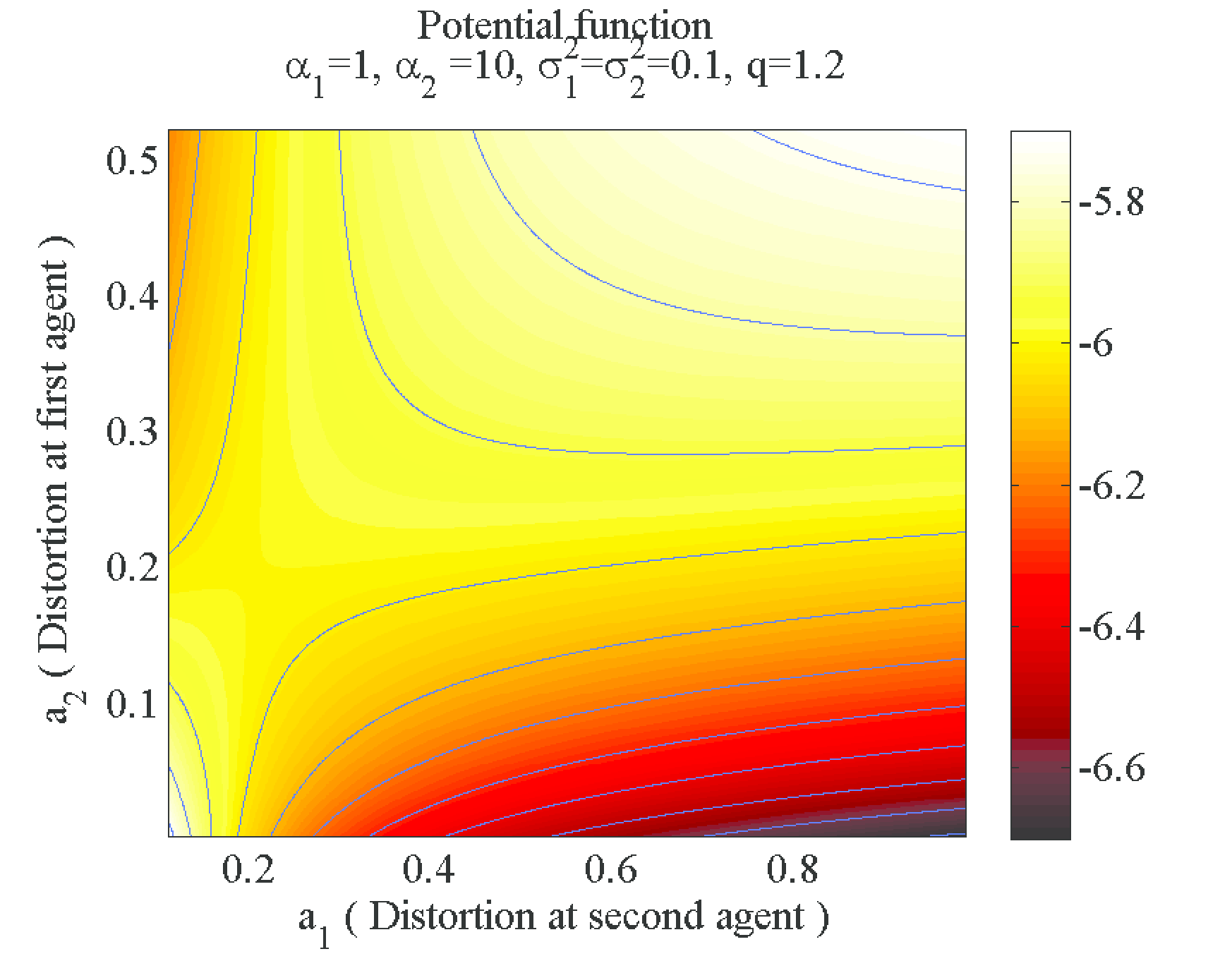}
\hspace{0.01\columnwidth}
\includegraphics[width=.9\columnwidth]{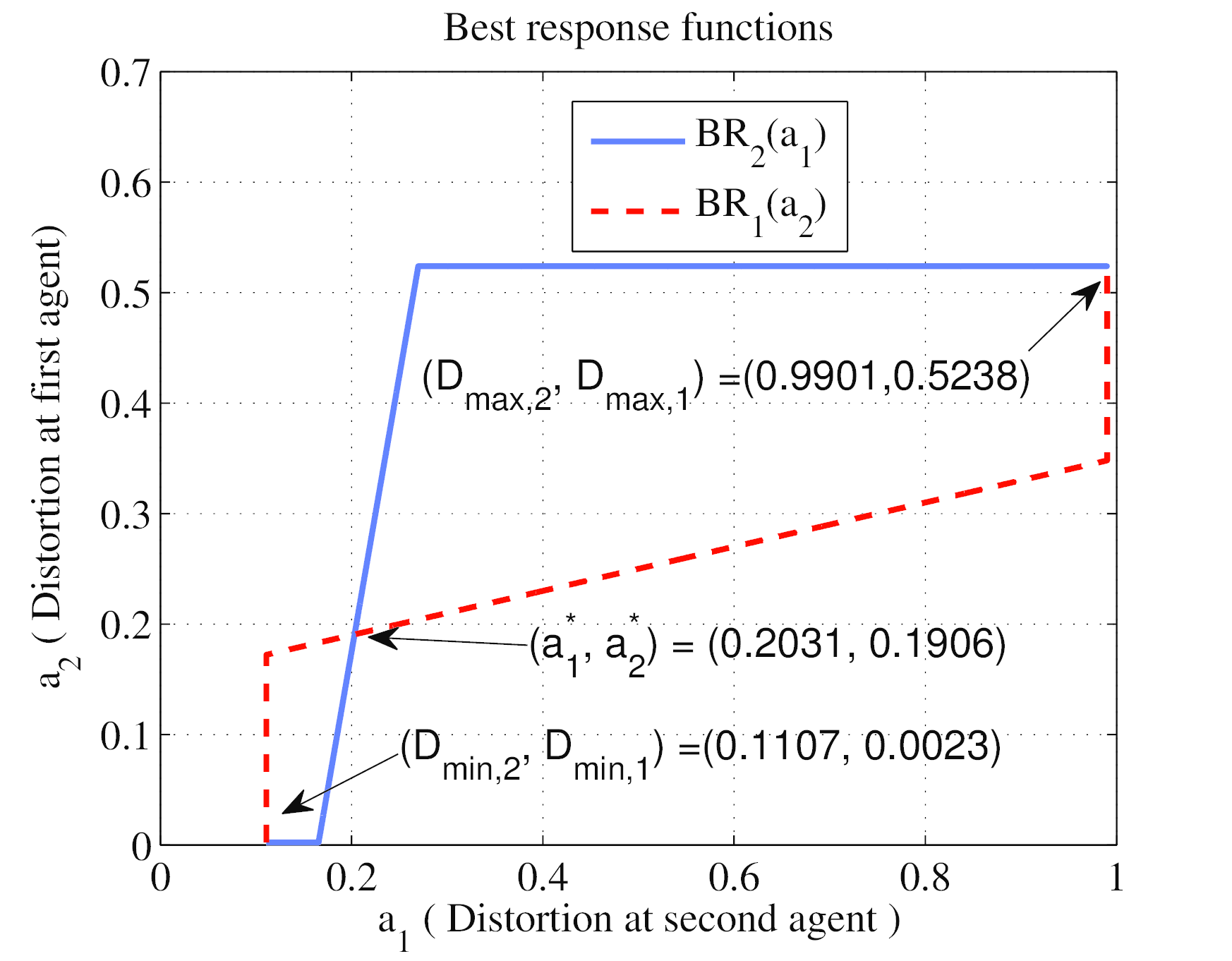} 
\caption{{\protect\scriptsize {Water-levels of the potential (up) and
BRs (down) as functions of $(a_{1},a_{2}) \in(D_{\min,2}, \overline{D}_{2}]
\times(D_{min,1}, \overline{D}_{1}]$. The potential has two local maxima
$(D_{\min,2}, D_{\min,1})$, $(\overline{D}_{2}, \overline{D}_{1})$ and one
saddle point $(a_{1}^{*}, a_{2}^{*})$. The saddle point is a NE not
asymptotically stable whereas the other two are asymptotically stable NE.}}}%
\label{fig:fig1_potential_3NE}%
\end{figure}
\end{center}

The case in which the game has three NEs is illustrated in Fig.
\ref{fig:fig1_potential_3NE} for the scenario: $\alpha=1$, $\alpha_2=10$,
$\sigma_{1}^{2} = \sigma_{2}^{2} = 0.1$ and $q=1.2$. The solutions are $(a_{1}^{\mathrm{NE}}, a_{2}^{\mathrm{NE}}) \in\left\{
(D_{\min,2}, D_{\min,1}), (\overline{D}_{2}, \overline{D}_{1}), (a_{1}^{*},
a_{2}^{*})\right\}  \equiv$ $\left\{  (0.1107, 0.0023), (0.9901,
0.5238),(0.2031, 0.1906)\right\}$. 

Analyzing the plot of the BR functions, we can observe that the intersection point $(a_{1}^{*},a_{2}^{*})$ is not asymptotically stable: Assume that a small perturbation moves the agents away from this point. By iterating the best responses, the agents get further away and converge to one of the other NEs. The initial perturbation determines which of the two NE - that are asymptotically stable - will be chosen. 

\begin{center}
\begin{figure}[h!]
\includegraphics[width=.9\columnwidth]{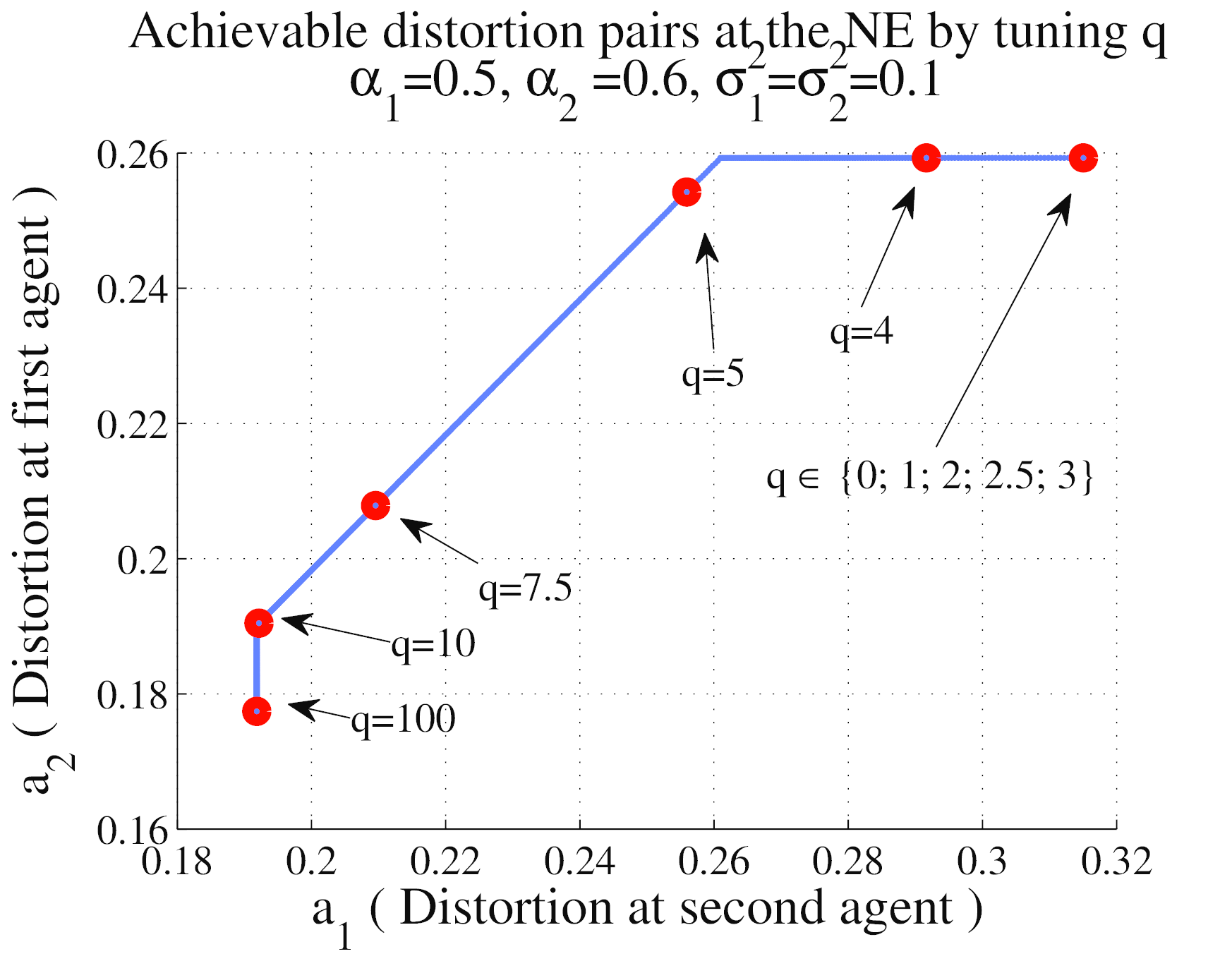}
\hspace{0.01\columnwidth}
\includegraphics[width=.9\columnwidth]{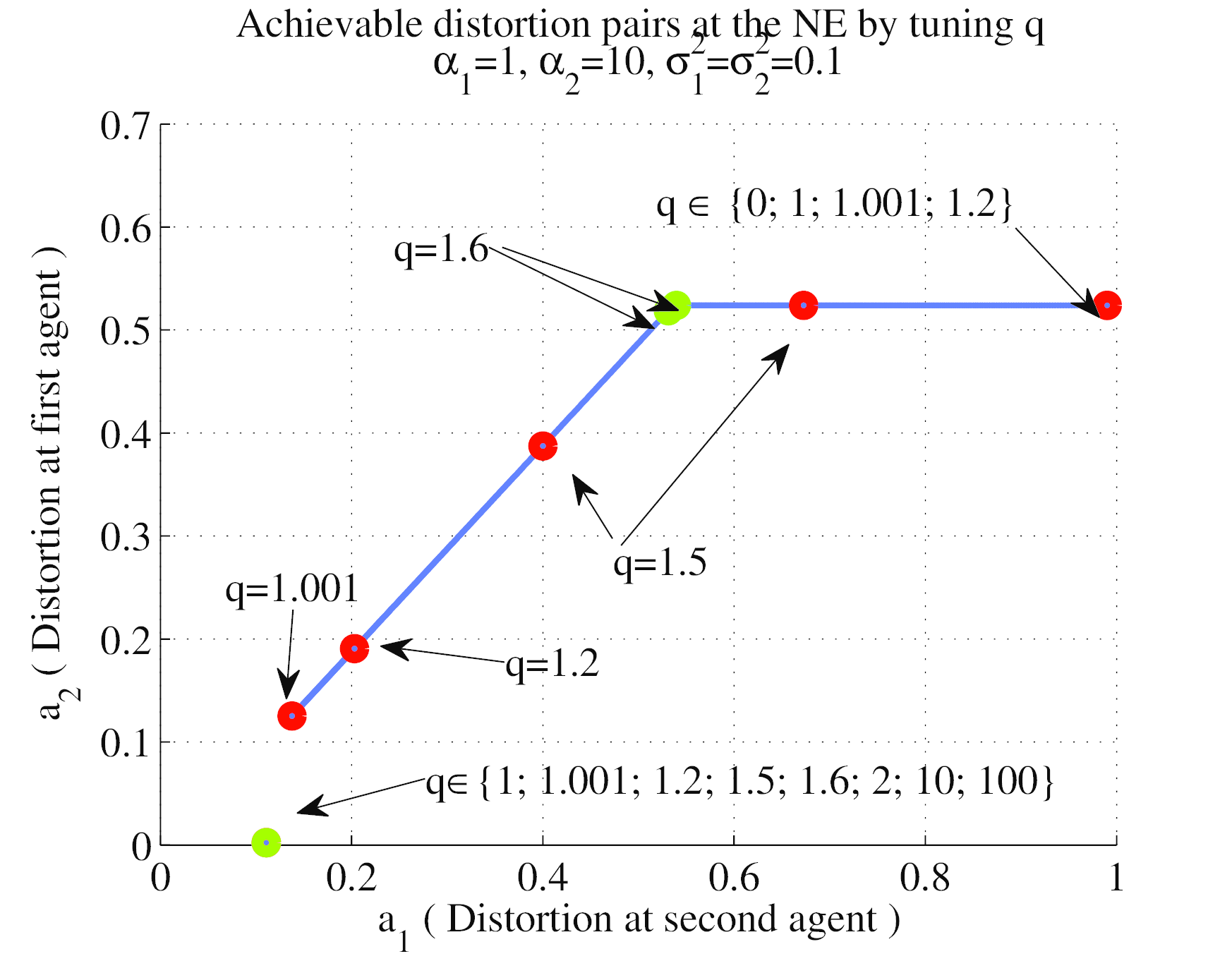} 
\caption{{\protect\scriptsize {Achievable distortion pairs at the NE obtained by tuning $q \in [0,100]$ for the scenarios of Fig. \ref{fig:fig1_potential_1NE} and Fig. \ref{fig:fig1_potential_3NE}. Not all distortions pairs can be achieved at the NE of the common-goal game. The distortion pairs that are achieved are the points which correspond to either the local maxima or saddle points of the system-wide objective function. }}}%
\label{fig:fig1_distortion_equil}%
\end{figure}
\end{center}

Fig. \ref{fig:fig1_distortion_equil} illustrates the NEs depending on the parameter $q \in [0, 100]$ tuned by the central controller. Both scenarios of Fig. \ref{fig:fig1_potential_1NE} and \ref{fig:fig1_potential_3NE} are considered.  

By choosing small values of $q$, the central controller prefers large distortions and small leakage tuples; privacy is enforced in the network. Larger values of $q$ result in opposite tuples (small distortions and large leakage tuples); cooperation is enabled among selfish agents. In the case of three NEs, the discontinuity at $q=1$ can be explained by the change in the BR functions; if $q>1$ they are continuous and piecewise affine; if $q \leq 1$ they are discontinuous Heaviside-type of functions (as seen in Sec. \ref{subsec:equal})).

We also remark that not all information-theoretic distortion-leakage tuples are achieved at the NE. Only the local maximizers or saddle points of the overall network-wide payoff function are NE and these tradeoff tuples depend on the system parameters. To achieve different tuples at the NE, other objective functions have to be considered (e.g., the sum of agents' individual payoff functions in  (\ref{eq:payoff_j_init})).

\vspace{-0.9in}

\section{Discounted repeated games}

\label{sec:repeated_game}


In large distributed networks, the need for continual monitoring makes repeated interactions among agents inevitable: The control of the electric power network depends on the state estimation performed periodically by distributed entities that interact with each other over and over. Such a repeated interaction may build trust among agents leading to sustained information exchange. 

As opposed to the previous section, we do not assume the presence of a central controller. Rather, we exploit the repetition aspect to achieve non-trivial distortion-leakage tuples naturally without economic incentives.

\paragraph*{One-shot game and pricing}

We start with a brief overview of the non-cooperative game introduced in \cite{belmega-isccsp-2012}. Consider the tuple $\mathcal{G}=(\mathcal{P}%
,\{\mathcal{A}_{j}\}_{j\in\mathcal{P}},\{u_{j}\}_{j\in\mathcal{P}})$, where
the set of players and their action sets are identical to the game
described in Sec. \ref{sec:potential_game}. The difference lies in the individual payoff functions: $u_{j}, \ \forall j\in \mathcal{P}$ which measures the satisfaction of agent $j$ and depends on his own action choice but also on the others' choices. As opposed to the common-goal game, each agent cares only of his own leakage of information and state estimation fidelity. Thus,
the payoff function of agent $j$, $u_{j}:\mathcal{A}_{j}\times\mathcal{A}%
_{i}\rightarrow\mathbb{R}$, is given by
\begin{equation}
u_{j}(a_{j},a_{i})= - L_{j}(a_{j})+ \frac{q_j}{2} \log\left(
\frac{\overline{D}_{j}}{a_{i}}\right)  .\label{eq:payoff_j_init}%
\end{equation}
The second term represents the information rate of the data received from the other agent depending on $a_{i}=D_{j}$, i.e., the distortion of agent $j$. The weight $q_j = \hat{w}_j/\tilde{w}_j$ is the ratio between the emphasis on leakage vs. state estimation fidelity of agent $j$.

Maximizing the utility in
(\ref{eq:payoff_j_init}) w.r.t. $a_j$ is equivalent minimizing only the first term: the leakage of information. Indeed, the second term is a result of the data shared by the other agent $i$, and hence, not in control of agent $j$. The game simplifies into two simple decoupled optimization problems; each agent chooses to stay silent (minimizing its leakage of information). The only rational outcome is the \emph{maximum distortion - minimum leakage} extreme for both agents.

\begin{remark}
The one-shot game $\mathcal{G}$ is somewhat similar to the
classical prisoners' dilemma \cite{tirole-book-1991} (which is a discrete game as opposed to our continuous game): each agent has a strictly dominant strategy\footnote{A strictly dominant strategy is an action that is the best choice of an agent independent from the others' choices.} which is that of \emph{not sharing any data (beyond the minimum requirement)}.
\end{remark}

In \cite{belmega-isccsp-2012}, we show that any tuple in the information-theoretic region is achievable provided the agents are appropriately rewarded. The modified payoff functions which include the pricing are:
\begin{equation}
\tilde{u}_{j}(a_{j},a_{i})=u_{j}(a_{j},a_{i})+\frac{p_{j}}{2}\log\left(\frac{\overline{D}_{i}%
}{a_{j}}\right).
\end{equation}

The drawback of such pricing techniques - that rewards an agent proportionality to his data sharing rate - is the implicit presence of a mediator (central controller or self-regulating market) which can manipulate the outcome by tuning the prices $p_{j}>0$. In the following, we show that repetition enables cooperation among selfish agents - without any centralized interference.


We assume that the agents interact with each other multiple times under the same conditions, i.e., they play the same non-cooperative game $\mathcal{G}$ repeatedly. The total number of rounds is denoted by $T\geq 1$. Two cases are distinguished in function of the available knowledge of $T$: (i) perfect knowledge of $T$ - both agents know in advance when their interaction ends; and (ii)
imperfect or statistical knowledge of $T$ - the agents do not know the precise ending of their interaction. 

\textit{In both cases, we study the possibility
of enabling and sustaining cooperation by allowing the agents to make only
credible commitments, i.e., commitments on which they have incentives to
follow through}. The equilibrium concept we investigate here is a refinement of the Nash equilibrium, i.e., \textit{subgame perfect equilibrium}, defined in the sequel. 

\subsection{Strategies, Payoffs and Subgame Perfect Equilibria}

We introduce some useful notation and definitions. These tools are necessary for a clear understanding of the solutions arising in repeated games. 

We assume that the game $\mathcal{G}$
described above is played several times. Repeated games
differ from one-shot games by allowing players to observe the history of the
game and condition their current play on past actions. The history at the end
of stage $t\geq1$ is denoted by $h^{(t+1)}=(a^{(1)},\hdots,a^{(t)})$, where
$a^{(\tau)}=(a_{1}^{(\tau)},a_{2}^{(\tau)})$ represents the agents' play or
action profile at stage $\tau$. The set of all possible histories at the end
of stage $t$ is denoted by $\mathcal{H}^{(t+1)}$ such that $\mathcal{H}^{(1)}$
denotes the void set. We can now formally define a repeated game.

\begin{definition}
A repeated game is a sequence of non-cooperative games given by the
tuple \newline$\mathcal{G}_{R}^{\left(  T\right)  }=(\mathcal{P}%
,\{\mathcal{S}_{j}\}_{j\in\mathcal{P}},\{v_{j}\}_{j\in\mathcal{P}},T)$, where
$\mathcal{P}\triangleq\{1,2\}$ is the set of \emph{players} (the two agents);
$\mathcal{S}_{j}$ is the strategy set of agent $j$; and $v_{j}$ is the payoff
function which measures the satisfaction of agent $j$ for any strategy
profile. 
\end{definition}

As opposed to the one-shot game, we have to make a clear distinction
between an action - the choice of an agent at a specific moment (or stage of the game) - and a strategy that describes the agents' behavior for the whole
duration of the game. A strategy of an agent is a contingent plan devising his play at each stage $t$ and
for any possible history $h^{(t)}$; more precisely it is defined as follows.

\begin{definition}
\label{def:strategy} A pure strategy for player $j$, $s_{j}$, is a
sequence of causal functions $\{ s_{j}^{(t)}\} _{1\leq t\leq T}$ such that
$s_{j}^{\left(  t\right)  }:\mathcal{H}^{\left(  t\right)  }\rightarrow
[D_{\min,i},\overline{D}_{i}]$, and $s_{j}^{\left(  t\right)  }\left(
h^{(t)}\right)  = a_{j}^{\left(  t\right)  }\in[D_{\min,i},\overline{D}_{i}]$.
\end{definition}

The set of strategies, denoted by $\mathcal{S}_{j}$, is the
set of all possible sequences of functions given in Definition
\ref{def:strategy}, such that, at each stage of the game, every possible
history of play $h^{(t)}$ is mapped into a specific action in $\mathcal{A}%
_{j}$ to be chosen at this stage.

In repeated games, the agents wish to maximize their averaged payoffs over the entire game horizon. We assume that agents discount future payoffs: present payoffs are more
important than future promises.

\begin{definition}
The discounted payoff function of player $j$ given a joint strategy $s=(s_{1}%
,s_{2})$ is given by
\begin{equation}
v_{j}(s)=(1-\rho_j)\sum_{t=1}^{T}\rho_j^{t-1}u_{j}(a^{(t)}%
),\label{eq:discounted_payoffs}%
\end{equation}
where $a^{(t)}$ is the action profile induced by the joint strategy $s$,
$u_{j}(\cdot)$ is the payoff function in (\ref{eq:payoff_j_init}), $\rho_j
\in(0,1)$ is the discount factor of player $j$. 
\end{definition}

The Nash equilibrium concept for repeated games is defined similarly to
the one-shot games (any strategy profile that is stable to
unilateral deviations). Some of the Nash equilibria of the repeated games may rely on \emph{empty threats} \cite{tirole-book-1991} of
suboptimal play at histories that are not expected to occur (under the players' rationality assumption). Thus, we focus on a subset
of Nash equilibria that allow players to make only commitments they have incentives to follow through: \emph{the subgame perfect equilibria}.

Before defining this concept, we have to define subgames. Given any history $h^{(t)}\in\mathcal{H}^{(t)}$, the game from stage $t$ onwards, is a subgame denoted by $\mathcal{G}_{R}(h^{(t)})$. The final history for this subgame is denoted by
$h^{(T+1)}=(h^{(t)},a^{(t)},\hdots,a^{(T)})$. The strategies and payoffs are
functions of the possible histories consistent with $h^{(t)}$. Any
strategy profile $s$ of the whole game induces a strategy $s|h^{(t)}$ on any
subgame $\mathcal{G}_{R}(h^{(t)})$ such that for all $j$, $s_{j}|h^{(t)}$ is
the restriction of $s_{j}$ to the histories consistent with $h^{(t)}$.

\begin{definition}
A subgame perfect equilibrium, $s^{*} = (s_{1}^{*}, s_{2}^{*})$, is a
strategy profile (in a repeated game with observed history) such that, for
any stage and any history $h^{(t)} \in\mathcal{H}^{(t)}$, the restriction $s^{*}|h^{(t)}$ is a Nash equilibrium for the subgame $\mathcal{G}_{R}(h^{(t)})$.
\end{definition}

This equilibrium concept is a refinement of the NE because it is required to be a NE in every possible subgame aside from the entire history game. We analyze this solution concept for two different cases in function of the available knowledge of the end stage: perfect knowledge and imperfect or statistical knowledge of $T$.

\vspace{-0.1in}

\subsection{\label{subsec:finite-horizon-RG}Perfect knowledge of end stage}

We assume the agents know in advance the value of $T$,
i.e., when the game ends precisely. We show that
data-sharing beyond the minimum requirement cannot be enabled in this case. 

\begin{corollary}
\label{Th_finite_H} Assuming the agents know perfectly the value of $T$, the discounted repeated game $\mathcal{G}_{R}^{(T)}$ has a unique subgame perfect equilibrium $s^*$ described by  
\textquotedblleft no data sharing beyond the minimum requirement \textquotedblright\ at each stage of the
game and for both agents:
\begin{equation}
s_{j}^{(t),\ast}=\overline{D}_{i},\ \forall t\in\{1,\hdots,T\},\forall
j\in\mathcal{P}.
\end{equation}

\end{corollary}

The proof is omitted as it follows similarly to the repeated prisoners' dilemma (using an extension of the backward induction principle to dominance solvable games \cite{tirole-book-1991}). The key element is the strict dominance principle: a rational player will never choose an action that is strictly dominated. The same result remains true if the discounted payoffs are replaced with average payoffs, $v_{j} (s) = \frac{1}{T} \sum_{t=1}^{T} u_{j}(a^{(t)})$. Moreover, Theorem \ref{Th_finite_H} extends to a general class called \emph{dynamic games} in
which the system parameters ($\alpha_1^{(t)}$, $\alpha_2^{(t)} $, $\sigma_{1}%
^{(t)}$, $\sigma_{2}^{(t)}$) may vary at every stage of the game. The same reasoning holds since, at any stage of the game, the action corresponding to ``no data sharing beyond the minimum requirement'' is the strictly dominating one.

The only achieved distortion-leakage tuple is the \emph{maximum distortion-minimum leakage} - similarly to the one-shot game. The main reason why cooperation is not sustainable is that agents know precisely when their interaction ends. Next, we consider that the agents interact over an indeterminate period (they are unsure of the precise ending). 

\vspace{-0.1in}

\subsection{Imperfect knowledge of end stage}

\label{subsec:infinite-horizon-RG}

We assume here that the players do not know the value of $T$ (the end stage). The discount factor $\rho_j$ can be interpreted as the agent's belief (or probability) that the interaction goes on (see \cite{letreust-twc-2010} and references therein). The probability that the game stops at stage $t$ is then $(1-\rho_j)\rho_j^{t-1}$. The discounted payoff  \eqref{eq:discounted_payoffs} represents an expected or average utility. Thus, we assume that agent $j$ know $\rho_j$ which models its belief on the interaction continuing or not, at every stage (the probability that the game goes on).

The strategy of playing the one-shot NE at every stage is a subgame perfect equilibrium in this case as well. 

\begin{theorem}
\label{Th_inf_nocoop} Assuming imperfect knowledge of the end stage and that $D_{\min,j}>0$ for all $j\in\mathcal{P}$, in the discounted repeated game
$\mathcal{G}_{R}^{(\rho)} = (\mathcal{P}, \left\{  \mathcal{S}_{j} \right\}
_{j \in\mathcal{P}}, \left\{  {v}_{j} \right\}  _{j \in\mathcal{P}})$, the strategy ``do
not share any information beyond the minimum requirement'' at each stage of
the game and for both agents is a subgame perfect equilibrium, i.e. :
\begin{equation}
\label{eq:inf_no_coop}s_{j}^{(t),*} = \overline{D}_{i}, \ \forall t \geq1,
\forall j \in\mathcal{P}.
\end{equation}
\end{theorem}
The details of the proof are reported in Appendix  \ref{appendix:C}. Unlike the case of perfect knowledge of $T$, we show that this is not the only possible outcome and other distortion-leakage pairs can be achieved.

Inspired from the repeated prisoners' dilemma, our objective is to show that non-trivial exchange of information can be sustainable. Consider the action profiles
$(D_{2}^{*}, D_{1}^{*}) \in[D_{\min,2}, \overline{D}_{2}) \times[D_{\min,1},
\overline{D}_{1}) $ which perform strictly better than the one-shot NE for both agents:
\begin{equation}
\label{eq:po_conditions}\left\{
\begin{array}
[c]{lcl}%
u_{1}(D_{2}^{*}, D_{1}^{*}) & > & u_{1} (\overline{D}_{2}, \overline{D}_{1})\\
u_{2}(D_{1}^{*}, D_{2}^{*}) & > & u_{2} (\overline{D}_{1}, \overline{D}_{2}).
\end{array}
\right.
\end{equation}
Such tuples may be expected to represent long term contracts or agreements between rational agents. Other tuples will never be acceptable: By not sharing any data, an agent is guaranteed at least the one-shot NE payoff value. In the game theoretic literature, these utility pairs are also known as \emph{individually rational} payoffs  \cite{aumann-book-1992}. 

\begin{center}
\begin{figure}
\caption{{\protect\scriptsize {\ The set of all payoff pairs. The low-left corner is the one-shot Nash equilibrium. The darker area is the subset of payoff pairs strictly better than the one-shot NE
$q_{1}=q_{2}=5$.}}}%
\label{fig:new_payoffs}%
\includegraphics[width=0.9\columnwidth]{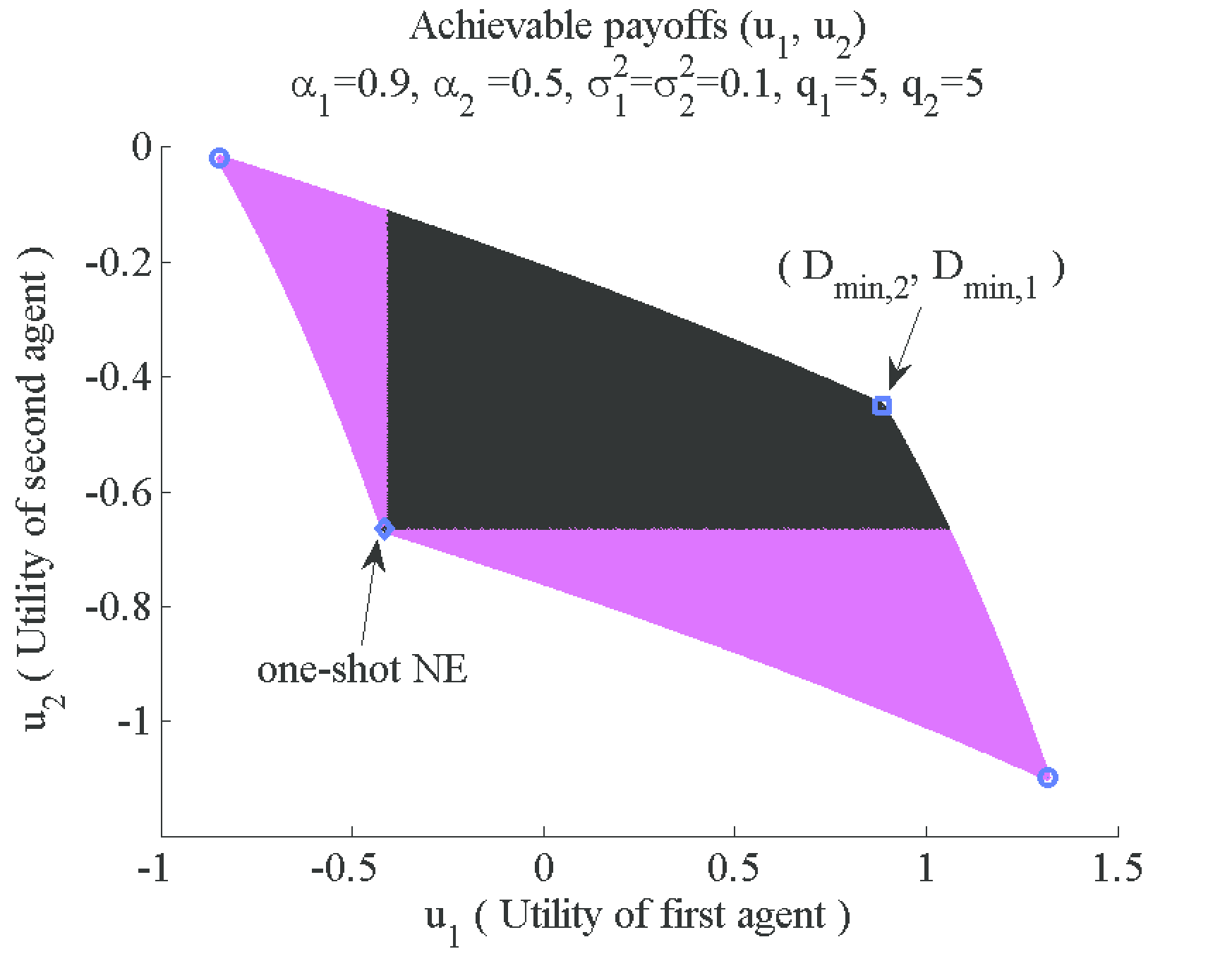} \vspace
{-0.2in}\end{figure}
\end{center}

These payoffs can be visualised in Fig. \ref{fig:new_payoffs} for the scenario: $\alpha_1= 0.9$, $\alpha_2= 0.5$, $\sigma_{1}^{2} = \sigma_{2}^{2} = 0.1$, $\overline{D}_{j} = D_{\min,j} + 0.5 (D_{\max,j} - D_{\min,j})$, $q_1=q_2=5$. The plotted area represents the set of all payoff pairs. The four corner points represent the four extremes: $(\overline{D}_{2}, \overline{D}_{1})$ (the low-left corner: the one-shot NE), $(D_{\min,2}, \overline{D}_{1})$ (the upper-left corner: the most advantageous for agent 2 - he shares nothing while agent 1 fully discloses his data), $(\overline{D}_{2}, D_{\min,1})$ (the low-right corner: the most advantageous for agent 1) and $(D_{\min,2}, D_{\min,1})$ (the upper-right corner: both agents fully disclose their data, maximizing their leakage). The darker area (in black) represents the subset of pairs satisfying \eqref{eq:po_conditions}. The lighter area (in magenta) represents the payoff pairs rejected by one or both rational players.
\vspace{-0.2in}
To gain more insight on these
achievable agreement points, we explicit the payoff
functions expressions in \eqref{eq:payoff_j_init}:
\begin{equation}
\label{eq:cond1}\left\{
\begin{array}
[c]{lcl}%
u_{j}(\overline{D}_{i}, \overline{D}_{j}) & = & - L_{j}%
(\overline{D}_{i})\\
u_{j}(D_{i}^{*}, D_{j}^{*}) & = & -  L_{j}({D}_{i}^{*}) + \frac{q_j}{2}
 \log\left( \frac{ \overline{D}_{j}}{D_{j}^{*}} \right)  .
\end{array}
\right.
\end{equation}
Data-sharing beyond the minimal requirement has two opposing effects: i) the leakage terms increase ($L_{j}%
(\overline{D}_{i}) < L_{j} (D_{i}^{*}), \ \forall\ D_{i}^{*} < \overline
{D}_{i}$); and ii) the estimation fidelity terms increase ($\log(\overline{D}_{j}/D_{j}^{*}) > 0,
\ \forall\ D_{j}^{*} < \overline{D}_{j}$).
Thus, the pairs $(D_{2}^{*},D_{1}^{*})$ represent the tuples which result in an increase of the state estimation fidelity that overcomes the loss
caused by the leakage for both agents. 

Intuitively, the greater the emphasis on the state estimation
terms, the larger the region of achievable agreement points is. We also
observe that the achievable distortion pairs satisfying the conditions in
\eqref{eq:po_conditions} must be relatively symmetric distortions pairs.
Otherwise said, both agents have to share their data for the agreement to be acceptable by both parties.

Unlike the one-shot game or the determined horizon repeated game (the agents have perfect knowledge of $T$), the commitment of
sharing data resulting in any distortion pair $(D_{2}^{*}, D_{1}^{*})$ is sustainable under some conditions on the discount factor. If the probability of the game stopping is small enough, then the commitment of playing $(D_{2}^{*}, D_{1}^{*})$ is credible  and, thus, sustainable to rational agents.

\begin{theorem}
\label{Th_inf_coop}  Assuming imperfect knowledge of the end stage in the discounted repeated game
$\mathcal{G}_{R} = (\mathcal{P}, \left\{  \mathcal{S}_{j} \right\}
_{j \in\mathcal{P}}, \left\{  {v}_{j} \right\}  _{j \in\mathcal{P}})$
and for any agreement profile $(D_{2}^{*},D_{1}^{*}) \in[D_{\min,2},
\overline{D}_{2}) \times[D_{\min,1}, \overline{D}_{1})$ that meets the
conditions \eqref{eq:po_conditions}, if the discount factors are bounded by: 
\begin{equation}
\label{eq:new_coop_condition}1 > \rho_j>  \frac{2[ L_{j}(D_{i}^{*}) - L_{j}(\overline{D}_{i})
]}{q_{j} \log\left(  \overline{D}_{j}/D_{j}^{*}\right)  },
\end{equation}
and $D_{\min,j}>0$ for all $j \in\mathcal{P}$, then the following strategy is
a subgame perfect equilibrium: For all $j$, ``agent $j$ shares data at the
agreement point $D_{i}^{*}$ in the first stage and continues to share data at
this agreement point if and as long as the other player $i$ shares data at the
agreement point $D_{j}^{*}$. If any player has ever defected from the
agreement point, then the players do not cooperate beyond the minimum
requirement from this stage on.'' 
\end{theorem}

A detailed proof is given in Appendix \ref{appendix:B}. This theorem assesses that both agents can achieve better
distortion levels than the one-shot NE naturally, without the interference of a central authority or economic incentives. The optimal strategy is a \emph{tit-for-tat} type of policy: \emph{Each agent fulfils his part of the agreement and shares data if and as long as the other party does the same.}

Any distortion pair $(D_{i}^{*}, D_{j}^{*})$ in
\eqref{eq:po_conditions} is achievable \emph{in the long term}, provided the discount factors are large enough. The lower bound in \eqref{eq:new_coop_condition} depends on the agents' emphasis on leakage vs. fidelity.
Larger emphasis on the leakage of information ($q_j\leq 1$) implies larger discount factors. Thus, smaller ending probability (or a longer expected interaction) is needed to sustain data sharing when agents are more sensitive to privacy concerns. 

This lower bound also depends on the 
specific agreement pair $(D_{2}^{*}, D_{1}^{*})$. It is again a compromise: Smaller distortion agreements imply larger leakages of information, thus, larger discount factors.  

In conclusion, the minimum expected length of the interaction needed to sustain an agreement depends on the agents' tradeoffs between the leakage of information and state estimation fidelity resulting from their data exchange.

Theorem \ref{Th_inf_coop} may be extended to the case in which the parameters
change at each stage of the game. However, the conditions on the discount
factor would be much stricter. A different approach should be investigated in such general dynamic games. This issue falls out the scope of
the present work and is left for future investigation.
\vspace{-0.1in}

\subsection{Numerical results}

\label{subsec:simus_RIH}

We focus on the scenario: $\alpha_1 = 0.9$, $\alpha_2= 0.5$, $\sigma_{1}^{2} = \sigma_{2}^{2} = 0.1$ and $\overline{D}_{j} = D_{\max,j}$ for $j \in\mathcal{P}$.  The minimum and maximum distortions are $D_{\min,1} = 0.3088$, $\overline{D}_{1} = 0.3926$, $D_{\min,2} = 0.2183$ and $\overline{D}_{2} = 0.2388$. For simplicity, we assume that both agents have the same belief on the end stage of the game, i.e.,  $\rho_1 = \rho_2 = \rho$.  

If the agents put an emphasis on leakage (e.g., $q_{1} = q_{2}=1$, $q_{1}=1, q_{2}=2$ or $q_{1}=2, q_{2}=1$, there is \emph{no distortion pair $(D_{2}^{*}, D_{1}^{*})$} that strictly improves both players' payoffs compared to the one-shot NE $(\overline{D}_{2},
\overline{D}_{1})$. This means that the improvement in an agent's estimation
fidelity from the data shared by the other agent is overcome by the loss of privacy incurred by the agreement point. 

\begin{center}
\begin{figure}[h!]
\includegraphics[width=.9\columnwidth]{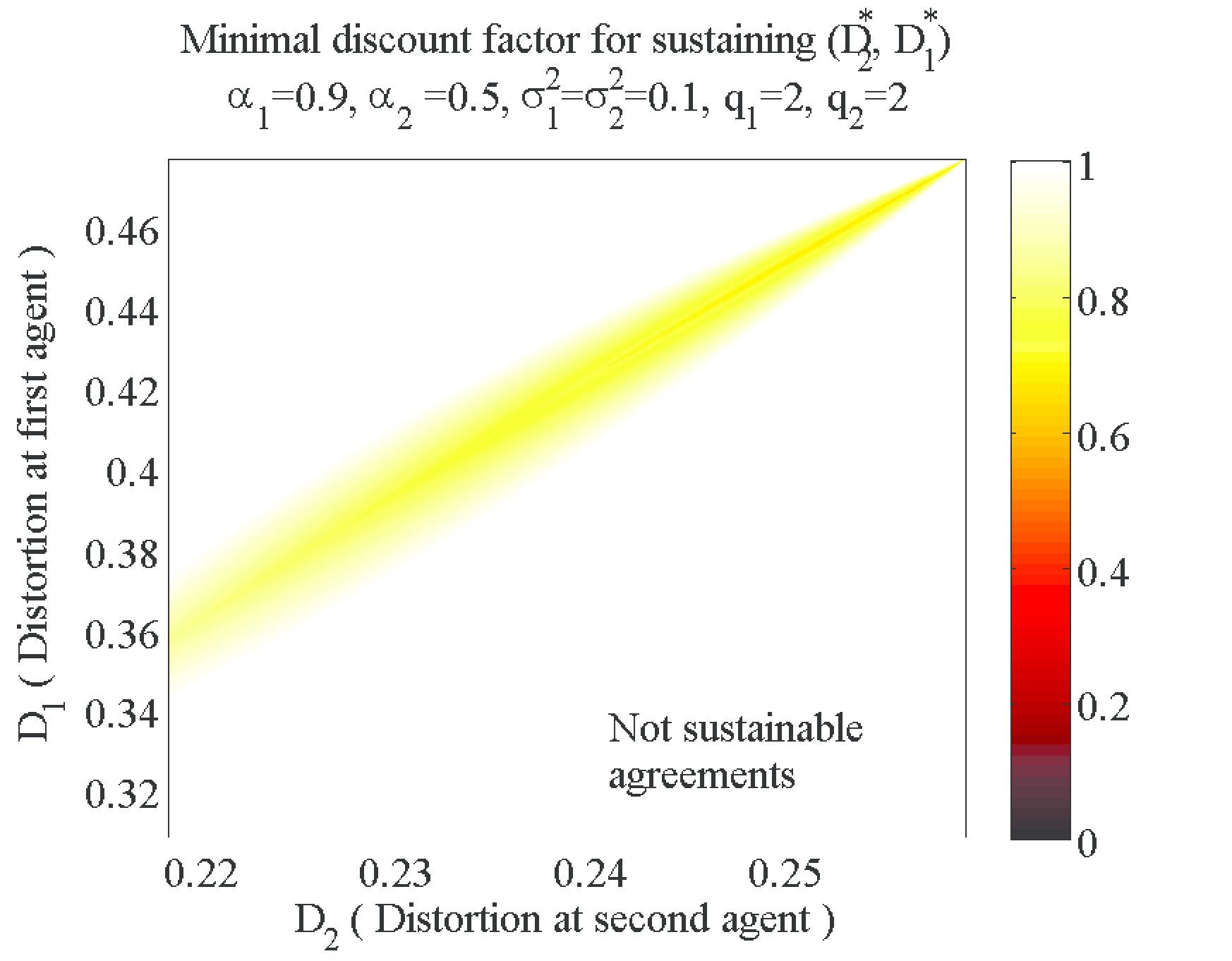}
\includegraphics[width =0.9\columnwidth]{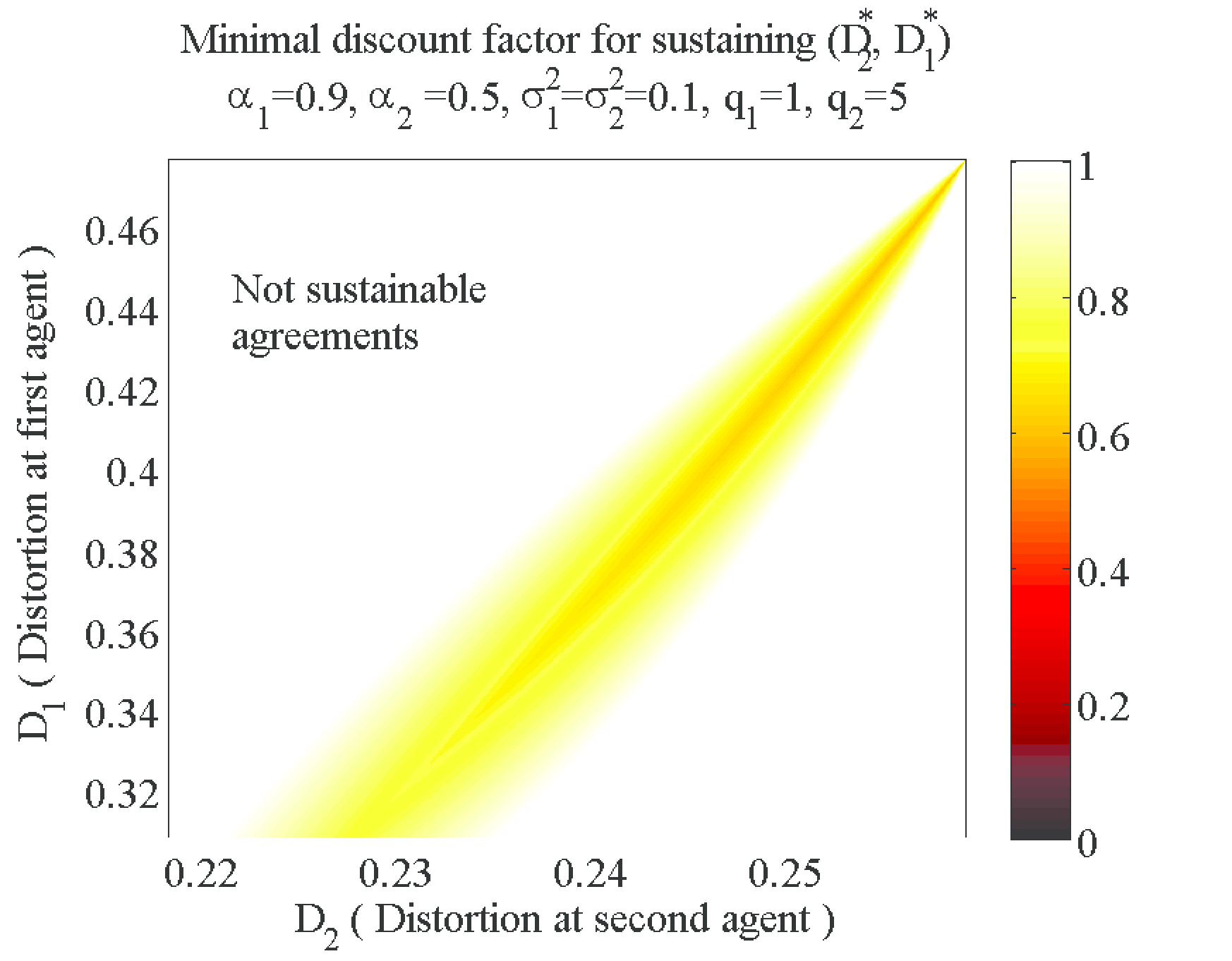}
\includegraphics[width=0.9\columnwidth]{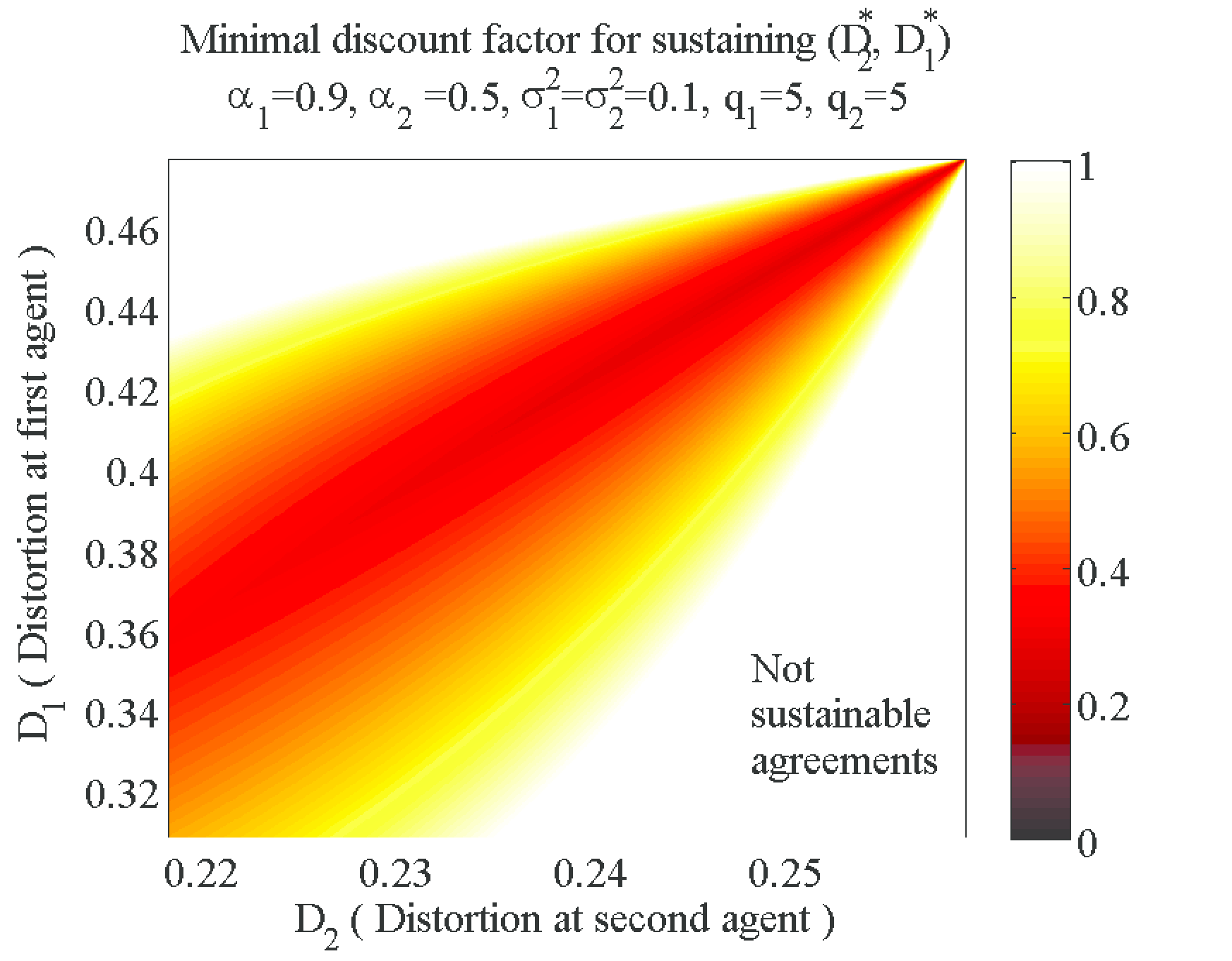}
\vspace{-0.2in}
\caption{{\protect\scriptsize {\ The subset of all possible
pairs $(D_{2}^{*},D_{1}^{*})$ that are achieved and the minimal discount factor $\rho$ needed to sustain them in the long-term interaction for the cases: i) $q_{1}=2, q_{2}=2$; ii) $q_{1}=1, q_{2}=5$; and iii) $q_{1}=5, q_{2}=5$.}}}%
\label{fig:fig1_repeated}%
\end{figure}
\end{center}

If the agents put more emphasis on their estimation fidelities, the region of agreements $(D_{2}^{*}, D_{1}^{*})$ becomes non-trivial. Figure  \ref{fig:fig1_repeated} illustrates this region in the cases: i) $q_{1}=2$, $q_{2}=2$; ii) $q_{1}=1$, $q_{2}=5$; and iii) $q_{1}=5$, $q_{2}=5$. The coloured region represents all the possible agreements sustainable in the long term, whereas the white region represents the distortion points that cannot be achieved. In all these figures, the upper-right corner represents the minimum cooperation requirement $(\overline{D}_{2}, \overline{D}_{1})$. 

Very asymmetric distortion pairs (the upper-left and lower-right regions) are not achievable in the long term; a rational user will only agree to fulfil equitable data-sharing agreements. In other words, either both players share information at a non-trivial rate or none of them does.

The higher the emphasis on state estimation fidelity, the larger the agreement region and lower the distortion levels achieved: The minimal distortion pair $(D_{\min,2}, D_{\min,1})$ is only sustainable in the third case ($q_1=q_2=5$) when the emphasis on the estimation fidelity is high enough for both agents.

We can observe a symmetry regarding the values of $\rho$ needed to sustain a given agreement pair. The fairer or more symmetric distortion pairs require a shorter expected game duration to be sustainable. The most unfair distortion pairs (the border points on the region of sustainable agreements) require the longest expected game duration; close to one probability of the game to continue. Beyond these edges, the difference between what an agent shares and what he receives in return is unacceptable, even in a long term interaction.

\vspace{-0.12in}

\section{Concluding Remarks}
\label{sec:clz}

Data sharing among physically interconnected nodes/agents of a network improves their local state estimations. When privacy also plays a role, enabling non-trivial data exchange often requires incentives. 

In a centralized setting, we show that the central controller can manipulate the data sharing policies of the agents by tuning a single parameter - depending on the emphasis between leakage vs. estimation fidelity. A whole range of outcomes can be chosen in between two extremes: both agents fully disclose their measurements (\emph{minimum distortion - maximum leakage}), and both agents stay silent (\emph{maximum distortion - minimum leakage}). 

If the network lacks a central controller and the agents are driven only by their individual agendas, we prove that non-trivial data sharing cannot be an outcome. Rational agents cannot trust each other in sharing data when the interaction takes place only once or in a finite number of rounds. However, if the agents interact repeatedly in the long term - over an undetermined number of rounds - then a whole region of outcomes is achieved depending on the agents' emphasis on leakage vs. state estimation fidelity. There is a symmetry in this achievable region: Rational agents agree only on \emph{tit-for-tat} data sharing policies. 


This results (long term repetition enables data exchange) follows from the underlying  assumption that agents can perfectly observe the past plays (the history of the game) and condition their present choices on these observations. In practice, this implies important signalling among the agents which has to be taken into account in future works. 

Although our work is focused on the case of two communicating agents, we make a first step in studying distributed solutions to competitive privacy problems in complex networks such as the electrical power network. Both our centralized and decentralized approaches use game theoretical tools which lead to developing distributed and scalable solutions. 

\vspace{-0.2in}

\appendices
\vspace{-0.2in}

\section{{\protect\small Proof of Theorem \ref{Theorem2}}}

\label{appendix:A}

Before providing the proof, we start by fully characterizing the set of NE. Three cases are distinguished depending on the $q$ parameter that determines the relative slopes of the BR functions. 
\begin{enumerate}
\item If $q > 2$, then there is a unique and asymptotically stable NE. If the intersection point of the affine functions $F_1(\cdot)$ and $F_2(\cdot)$ denoted by $(a_1^*, a_2^*)$ with 
\begin{equation}
\left\{ \begin{array}{lcl}
a_1^* & = & \frac{q}{1-(q-1)^2} \left( \frac{\delta_1}{\gamma_1} (q-1) + \frac{\delta_2}{\gamma_2} \right)\\
a_2^* & = & \frac{q}{1-(q-1)^2} \left( \frac{\delta_2}{\gamma_2} (q-1) + \frac{\delta_1}{\gamma_1} \right).
\end{array} \right.
\end{equation}
lies in the interior of $\Delta$, then it is the NE of the game. Otherwise, the NE lies on the border of $\Delta$.
\item If $q = 2$, then we have two different situations. If the condition $\delta_1/\gamma_1 + \delta_2/\gamma_2 \neq  0$ holds, then there is a unique and asymptotically stable NE lying on the border of $\Delta$. If on the contrary $\delta_1/\gamma_1 + \delta_2/\gamma_2 = 0$, then $F_i(a_j) \equiv F_j^{-1}(a_j)$. In this case, if this affine function intersects $\Delta$ non-trivially, then the game has an infinite number of NEs which are not asymptotically stable. Otherwise, the unique NE lies on the border and is asymptotically stable.
\item If $q < 2$, then there are two or three different NEs provided that the intersection point in (\ref{eq:intersection}) lies in the interior or on the border of $\Delta$: this intersection point is the only asymptotically unstable equilibrium. The other one or two NEs lie on the corners of $\Delta$, $(D_{\min,2}, D_{\min,1})$ and
$(\overline{D}_{2}, \overline{D}_{1})$). Otherwise, there is a unique NE which lies on the border of $\Delta$ and is asymptotically stable. 
\end{enumerate}

Intuitively, the scalar threshold equal to $2$ for the parameter $q$ comes
from the relative order among the two slopes of the BR functions. If $q=2$,
then the two slopes are identical and equal to one. In any other case, the
slopes of the two curves are different in the same axis system $\widehat{a_1 O a_2}$ (since
one of the two curves would have to be inverted). The relative slopes of the
two curves greatly influence their intersection points and, thus, the set of
NE.

The proof follows a similar approach as in \cite{belmega-eurasip-2010} for the
power allocation game over non-overlapping frequency bands in the interference
relay channel and assuming a zero-delay scalar amplify-and-forward relaying
protocol. We investigate the NEs of the game $\mathcal{G}_{\mathrm{sys}}$ when
$q>1$ and their asymptotic stability. A necessary and sufficient condition
that guarantees the asymptotic stability of a certain NE, say $(a_{1}%
^{\mathrm{NE}}, a_{2}^{\mathrm{NE}})$, is related to the relative slopes of
the BRs \cite{tirole-book-1991}:%
\begin{equation}
\label{eq:asym_stab_NE}\left|  \frac{\partial BR_{1}}{\partial a_{2}}(a_{2})
\frac{\partial BR_{2}}{\partial a_{1}} (a_{1}) \right|  < 1
\end{equation}
for all $(a_{1},a_{2})$ in an open neighbourhood of $(a_{1}^{\mathrm{NE}},
a_{2}^{\mathrm{NE}})$.
The analysis of the NE is based on the analysis of
intersection points of the two BR functions in \eqref{eq:br_fidelity}.

First, we analyze all the possible cases in which the intersection points
between the affine functions $F_{1}(\cdot)$ and $F_{2}(\cdot)$ are outside the
interval $\Delta$ or on the two corners: $(D_{\min,2}, \overline{D}_{1})$ or
$(\overline{D}_{2}, D_{\min,1})$. In these cases, the NE is unique and it lies
on the border of $\Delta$. These cases correspond to: (i) $F_{1}(\overline
{D}_{1}) \leq D_{\min,2} $ or $F_{2}(D_{\min,2}) \geq\overline{D}_{1}$, (ii)
$F_{1}(D_{\min,1}) \geq\overline{D}_{2}$ or $F_{2}(\overline{D}_{2})\leq
D_{\min,1}$ and the corresponding analysis will not be reported here as they
are tedious and similar to the next more interesting one. The more interesting case is when $F_{1}(D_{\min,1}) < \overline{D}_{2}$,
$F_{1}(\overline{D}_{1}) > D_{\min,2}$, $F_{2}(D_{\min,2}) < \overline{D}_{1}$
and $F_{2}(\overline{D}_{2})> D_{\min,1}$. This means that, if the curves
$F_{1}(\cdot)$ and $F_{2}(\cdot)$ intersect, the intersection point or points
lie in $\Delta$ and are NEs of the game under study. We have again three sub-cases:

\paragraph{If $q=2$} then the two functions $F_{1}(\cdot)$ and $F_{2}^{-1}(\cdot)$
have the same slope (equal to one) and thus they are parallel. 

\begin{itemize}
\item If $\delta_{1}/\gamma_{1} = - \delta_{2}/\gamma_{2}$, then the two
functions are the same. All the points on these curves that intersect $\Delta$
are NEs of the game. Therefore, we have an infinite number of NEs. The
asymptotic stability condition is not met because 
$$\left|  \frac{\partial BR_{1}}{\partial a_{2}}(a_{2})
\frac{\partial BR_{2}}{\partial a_{1}} (a_{1}) \right|  = 1,$$ for all
these NEs.
\item If $\delta_{1}/\gamma_{1} \neq- \delta_{2}/\gamma_{2}$, then the two BR
function intersect on the border of $\Delta$ in a unique asymptotically stable
point for which
$$\left|  \frac{\partial BR_{1}}{\partial a_{2}}(a_{2})
\frac{\partial BR_{2}}{\partial a_{1}} (a_{1}) \right|  =0.$$ 
\end{itemize}

\paragraph{If $q>2$} then the NE is unique and a detailed discussion follows
depending on the signs of the following inequalities: $F_{1}(D_{\min
,1})\lesseqgtr D_{\min,2}$, $F_{1}(\overline{D}_{1})\lesseqgtr\overline{D}%
_{2}$, $F_{2}(D_{\min,2}) \lesseqgtr D_{\min,1}$ and $F_{2}(\overline{D}_{2})
\lesseqgtr\overline{D}_{1}$ and also on the relative positions of the
intersection points between the two $F_{j}(\cdot)$ functions and the border of
$\Delta$. We will detail only one of these cases. \newline If $F_{1}%
(D_{\min,1})\geq D_{\min,2}$, $F_{1}(\overline{D}_{1}) \leq\overline{D}_{2} $,
$F_{2}(D_{\min,2}) \geq D_{\min,1}$ and $F_{2}(\overline{D}_{2}) \leq
\overline{D}_{1}$, then the two BR functions coincide on $\Delta$ with the two
functions $F_{j}(\cdot)$. The unique NE is given by the intersection point $(a_{1}^{*}, a_{2}^{*})$ of $F_{1}(\cdot)$ and $F_{2}(\cdot)$ such that
\vspace{-0.075in}
\begin{equation}%
\vspace{-0.075in}
\begin{array}
[c]{lcl}%
a_{1}^{*} & = & \frac{q}{1-(q-1)^{2}} \left(  \frac{\delta_{1}}{\gamma_{1}}
(q-1) + \frac{\delta_{2}}{\gamma_{2}} \right) \\
a_{2}^{*} & = & \frac{q}{1-(q-1)^{2}} \left(  \frac{\delta_{2}}{\gamma_{2}}
(q-1) + \frac{\delta_{1}}{\gamma_{1}} \right)  .
\end{array}
\end{equation}
It is easy to see that 
$$\left|  \frac{\partial BR_{1}}{\partial a_{2}}(a_{2})
\frac{\partial BR_{2}}{\partial a_{1}} (a_{1}) \right|  < 1$$ and, thus, the NE is
asymptotically stable.

\paragraph{If $q<2$} then the discussion follows similarly depending on the signs of the following inequalities: $F_{1}(D_{\min,1})\lesseqgtr D_{\min,2}$,
$F_{1}(\overline{D}_{1})\lesseqgtr\overline{D}_{2}$, $F_{2}(D_{\min,2})
\lesseqgtr D_{\min,1}$ and $F_{2}(\overline{D}_{2}) \lesseqgtr\overline{D}%
_{1}$ and also on the relative intersection points between the $F_{j}(\cdot)$
functions with the border of $\Delta$. 


\section{{\protect\small Proof of Theorem \ref{Th_inf_nocoop}}}

\label{appendix:C}

The backward induction argument is no longer valid since agents
do not know which stage is the final one. Instead, we apply the
\emph{one-stage-deviation principle} for discounted repeated games that are uniformly bounded in each stage
\cite{tirole-book-1991}. This principle states that a strategy profile
$s^{*}=(s_{1}^{*}, s_{2}^{*})$ is subgame perfect if and only if there is no
player $j$ and strategy $\hat{s}_{j}$ that agrees with $s^{*}_{j}$ except at a
single stage $\tau$ and history $h^{(\tau)}$, and such that $\hat{s}%
_{j}|h^{(\tau)}$ is a better response than $s^{*}_{j}|h^{(\tau)}$ in the
subgame $\mathcal{G}_{R}^{(\rho) }(h^{(\tau)})$.

First, we have to check the uniform boundedness condition on the stage
payoffs. Indeed, we can show that the stage payoffs in
\eqref{eq:payoff_j_init} are bounded as follows:
\begin{equation}
\label{eq:unif_bound}%
\begin{array}
[c]{lcl}%
|u_{j}(a_{j}^{(t)},a_{i}^{(t)})| & \leq &  L_{j}(a_{j}^{(t)}) +
\frac{q_j}{2} \log\left(  \overline{D}_{j}/a_{i}^{(t)} \right) \\
& \leq &(1 + {q}_{j}) \frac{1}{2} \log\left(  1/D_{\min,j} \right)  .
\end{array}
\nonumber
\end{equation}
Given that $D_{\min,j} < 1$, under the mild assumptions that $D_{\min,j} > 0$ and that $q_{j}$ is finite, the stage payoffs are uniformly bounded.

Second, we have to check whether unilateral deviation in a single stage from
the strategy in \eqref{eq:inf_no_coop} can be profitable. If not, then the
strategy is a subgame perfect equilibrium. Assume that player $j$ deviates at
time $\tau$ and history $h^{(\tau)}$ by choosing $\hat{s}_{j}^{(\tau)}
(h^{(\tau)}) = \hat{D}_{i} \in(D_{\min,i}, \overline{D}_{i})$ at stage $\tau$.
From then on, this strategy conforms to $s^{*}$, i.e., $\hat{s}_{j}^{(t)}
\equiv s_{j}^{*,(t)}$, for all $t>\tau$. This means that the leakage of
information of player $j$ will increase at stage $\tau$ and therefore its
payoff will decrease: $u_{j}(\hat{D}_{i}, \overline{D}_{j}) < u_{j}%
(\overline{D}_{i}, \overline{D}_{j})$. This implies directly that $v_{j}
(\hat{s}_{j}|h^{(\tau)}, s_{i}^{*}|h^{(\tau)}) < v_{j} (s_{j}^{*}|h^{(\tau)},
s_{i}^{*}|h^{(\tau)})$. Therefore, no agent has any interest in deviating at
any single stage and the plan defined in \eqref{eq:inf_no_coop} represents a
subgame perfect equilibrium.

As opposed to the case in which the perfect knowledge of $T$ is available, the discounted payoffs play a crucial role in the one-stage-deviation principle and, thus, this proof is not readily applicable in the case where a uniform average of the stage-payoffs is considered. Also, similarly to the case of perfect knowledge of $T$, this result extends to a general class of dynamic games in which the system parameters change at every stage
of the game.

\section{{\protect\small Proof of Theorem \ref{Th_inf_coop}}}
\vspace{-0.05in}
\label{appendix:B}
We use the one-stage deviation principle similarly to the proof of
Theorem \ref{Th_inf_nocoop}. Assume that no agent deviates in any subgame from
the agreement point. In this case, the discounted long-term payoff of player
$j$ is equal to $u_{j}(D_{i}^{*}, D_{j}^{*})$, i.e., the instantaneous payoff
achieved at the agreement point. If a player $j$ deviates at stage $\tau$ by
choosing $\hat{s}_{j}^{(\tau)} = {D}_{i} > D_{i}^{*}$ and then onwards
conforms to the strategy by choosing ${D}_{i}$, his discounted payoff is
\begin{equation}
\label{eq:inf_deviator}%
\begin{array}
[c]{l}%
(1-\rho_j) \displaystyle{\sum_{t=0}^{\tau-2}} \rho_j^{t} u_{j}(D_{i}^{*},
D_{j}^{*}) + (1-\rho_j)\rho_j^{\tau-1} u_{j}({D}_{i}, D_{j}^{*}) +\\
(1-\rho_j)\rho_j^{\tau} \displaystyle{\ \sum_{t=0}^{+\infty} } \rho^{t}
u_{j}(\overline{D}_{i}, \overline{D}_{j}) \ = \ u_{j}(D_{i}^{*}, D_{j}^{*})
\ -\\
\rho_j^{\tau-1} \left[  u_{j}(D_{i}^{*}, D_{j}^{*}) - u_{j} ({D}_{i}, D_{j}^{*})
\ + \right. \\
\\
\left.  \rho_j\left(  u_{j}({D}_{i}, D_{j}^{*}) - u_{j}(\overline{D}_{i},
\overline{D}_{j}) \right)  \right]  .
\end{array}
\end{equation}
Notice that $u_{j} ({D}_{i}, D_{j}^{*})- u_{j}(D_{i}^{*}, D_{j}^{*}) > 0$ (by
sharing more information, the leakage term for player $j$ increases and his
payoff decreases), and $u_{j}({D}_{i}, D_{j}^{*}) - u_{j}(\overline{D}_{i},
\overline{D}_{j}) >0$ (from the previous observation and condition
\eqref{eq:po_conditions}) for any ${D}_{i} > D_{i}^{*}$ . Under the following
sufficient condition on the discount factor: \vspace{-0.15in}
\begin{equation}
\label{eq:coop_condition}1 > \rho_j> \displaystyle{\ \max_{
\ {D}_{i} \in(D_{i}^{*}, \overline{D}_{i}]} \left\{  \frac{u_{j}( {D}_{i},
D_{j}^{*}) - u_{j}(D_{i}^{*}, D_{j}^{*})}{u_{j}( {D}_{i}, D_{j}^{*}) -
u_{j}(\overline{D}_{i}, \overline{D}_{j})} \right\}  },
\end{equation}
this discounted payoff for the deviator in \eqref{eq:inf_deviator} is less
than the payoff of no deviation $u_{j}(D_{i}^{*}, D_{j}^{*})$. 

Now, let us assume that a deviation has occurred. At stage $\tau$ and for any
history $h^{(\tau)}$ after this deviation, if player $j$ were to deviate from
the prescribed strategy and choose $\hat{s}_{j}^{(\tau)} = {D}_{i} <
\overline{D}_{i}$ and then conform from this stage onwards, its leakage term
would increase and its payoff in stage $\tau$ would be strictly less than if
it had not deviated. Thus, no player has any incentive to deviate at any
single stage of the game and for any history of play and the strategy
described in this theorem is a subgame perfect equilibrium for the discounted repeated game in which the end stage of the game is not known.

To complete the proof, we have to show that the sufficient condition in
\eqref{eq:coop_condition} is equivalent to the one in
\eqref{eq:new_coop_condition}. First, from \eqref{Th_L1} we observe that the
leakage function $L_{j}(D_{i})$ is strictly decreasing with $D_{i}$ (the
smaller distortion at agent $i$ the bigger the leakage term), and, thus, we
have $\frac{\mathrm{d}L_{j}}{\mathrm{d} D_{i}} < 0$. Second, by replacing the
payoff functions expressions in \eqref{eq:payoff_j_init} we have:

\begin{align}
{\small \frac{u_{j}( {D}_{i}, D_{j}^{*}) - u_{j}(D_{i}^{*},
D_{j}^{*})}{u_{j}( {D}_{i}, D_{j}^{*}) - u_{j}(\overline{D}_{i}, \overline
{D}_{j})} =}\nonumber\\
\frac{ [L_{j}(D_{i}^{*}) - L_{j}(D_{i})]}{
[L_{j}(\overline{D}_{i}) - L_{j}(D_{i})] + \frac{q_j}{2} \log\left(
\overline{D}_{j}/D_{j}^{*}\right)  }.\nonumber
\end{align}
 
 We compute the
derivative of the right-side term w.r.t. $D_{i}$ and obtain: \vspace{-0.1in}
\begin{align}
 {\small \frac{\mathrm{d}}{\mathrm{d}D_{i}} \ 
\frac{L_{j}(D_{i}^{*}) - L_{j}(D_{i})}{L_{j}(\overline{D}_{i}) - L_{j}(D_{i}) + \frac{q_{j}}{2} \log\left(
\frac{\overline{D}_{j}}{D_{j}^{*}}\right)  }   =}\nonumber\\
-  \frac{ L_{j}(\overline{D}_{i}) - L_{j}%
(D_{i}^{*}) + \frac{q_{j}}{2} \log\left(  \frac{\overline{D}_{j}}{D_{j}^{*}}\right)  }{
\left[  L_{j}(\overline{D}_{i}) - L_{j}(D_{i})] + \frac{q_{j}}{2}
\log\left(  \frac{\overline{D}_{j}}{D_{j}^{*}}\right)  \right]  ^{2}} \left(
\frac{\mathrm{d} L_{j}}{ \mathrm{d} D_{i}} \right)  . \label{eq:deriv}%
\end{align}
From the fact that $\frac{\mathrm{d}L_{j}}{\mathrm{d} D_{i}} < 0$, equations
\eqref{eq:po_conditions} and \eqref{eq:cond1}, we obtain that the derivative
in \eqref{eq:deriv} is strictly positive, and, thus
\begin{align}
{\small \allowdisplaybreaks \displaystyle{\ \max_{{D}_{i}
\in(D_{i}^{*}, \overline{D}_{i}]} \left\{  \frac{u_{j}( {D}_{i}, D_{j}^{*}) -
u_{j}(D_{i}^{*}, D_{j}^{*})}{u_{j}( {D}_{i}, D_{j}^{*}) - u_{j}(\overline
{D}_{i}, \overline{D}_{j})} \right\}  } =} \nonumber\\
\displaystyle{ 2 \frac{
L_{j}(D_{i}^{*}) - L_{j}(\overline{D}_{i}) }{q_{j} \log\left(
\frac{\overline{D}_{j}}{D_{j}^{*}}\right)  }  }.\nonumber
\end{align}

\bibliographystyle{IEEEtran}
\bibliography{IEEEabrv,biblio_07apr2014,biblio_09aug2013}

\end{document}